\documentclass[aps,prd,twocolumn,superscriptaddress,nofootinbib]{revtex4-1}

\usepackage{graphicx,color,amsmath,amssymb}
\usepackage{slashed}
\usepackage{soul}
\usepackage{aastex}
\usepackage{amsmath}
\usepackage{braket}
\usepackage[utf8]{inputenc}
\usepackage{pifont}
\usepackage{array}
\newcommand{\nocontentsline}[3]{}
\newcommand{\epar}{E_\parallel}

\newcommand{\rpc}{r_{\text{pc}}}

\newcommand{\gammamax}{\gamma_{\text{max}}}
\newcommand{\tacc}{t_\text{acc}}

\newcommand{\gagg}{g_{a\gamma\gamma}}
\newcommand{\tpmcreate}{$t^\pm_\text{create}$}
\newcommand{\zpmcreate}{$z^\pm_\text{create}$}
\newcommand{\tgcreate}{$t^\gamma_\text{create}$}
\newcommand{\zgcreate}{$z^\gamma_\text{create}$}

\usepackage[export]{adjustbox}
\usepackage{tikz}
\usepackage{tkz-euclide}
\usetikzlibrary{decorations.pathmorphing}   
\tikzset{
    v/.style={decorate, decoration={snake, segment length=3mm, amplitude=0.75mm}, draw},
    f/.style={draw=black, postaction={decorate},
        decoration={markings,mark=at position .6 with {\arrow[very thick]{latex}}}},
    fb/.style={draw=black, postaction={decorate},
        decoration={markings,mark=at position .4 with {\arrowreversed[very thick]{latex}}}},
    fnar/.style={draw=black},
    g/.style={decorate, draw=black,
        decoration={coil,amplitude=3pt, segment length=3.5pt}},
    s/.style={dashed,draw=black, postaction={decorate},
        decoration={markings,mark=at position .55 with {\arrow[very thick]{latex}}}},
    sb/.style={dashed,draw=black, postaction={decorate},
        decoration={markings,mark=at position .55 with {\arrowreversed[draw=black,very thick]{latex}}}},
    snar/.style={dashed,draw=black,line width =1.25pt},
}

\usepackage{hyperref} 
\usepackage{amsmath}
\hypersetup{
    colorlinks=true,
    citecolor=[rgb]{.1, .7, .6},
    linkcolor=[rgb]{.2, .55, .95},
    filecolor=magenta,
    urlcolor=[rgb]{.1, .7, .6},
}

\newcommand{\eV}{{\, {\rm eV}}}

\newcommand{\MHz}{{\, {\rm MHz}}}
\newcommand{\GHz}{{\, {\rm GHz}}}

\newcommand{\rns}{R_{\rm NS}}

\usepackage{tikz}
\newcommand{\tikzxmark}{%
\tikz[scale=0.23] {
    \draw[line width=0.7,line cap=round] (0,0) to [bend left=6] (1,1);
    \draw[line width=0.7,line cap=round] (0.2,0.95) to [bend right=3] (0.8,0.05);
}}

\newcommand{\ie}{{\it i.e.~}}  \newcommand{\eg}{{\it e.g.~}}
\definecolor{mypurple}{RGB}{164,64,214}

\definecolor{darkgreen}{rgb}{0.0, 0.5, 0.0}

\begin{document}

\title{Novel Constraints on Axions Produced in Pulsar Polar-Cap Cascades}
\author{Dion Noordhuis}
\thanks{These authors contributed equally to this manuscript.}
\affiliation{GRAPPA Institute, Institute for Theoretical Physics Amsterdam and Delta
Institute for Theoretical Physics,
University of Amsterdam, Science Park 904, 1098 XH Amsterdam, The Netherlands}
\author{Anirudh Prabhu}
\thanks{These authors contributed equally to this manuscript.}
\affiliation{Princeton Center for Theoretical Science, Princeton University, Princeton, NJ 08544, USA}
\affiliation{Stanford Institute for Theoretical Physics, Stanford University, Stanford, CA 94305, USA}
\author{Samuel J. Witte}
\thanks{These authors contributed equally to this manuscript.}
\affiliation{GRAPPA Institute, Institute for Theoretical Physics Amsterdam and Delta
Institute for Theoretical Physics,
University of Amsterdam, Science Park 904, 1098 XH Amsterdam, The Netherlands}
\author{Alexander Y. Chen}
\affiliation{Physics Department and McDonnell Center for the Space Sciences, Washington University, St. Louis, MO 63130, USA}
\author{F\'{a}bio Cruz}
\affiliation{GoLP/Instituto de Plasmas e Fus\~{a}o Nuclear,
Instituto Superior T\'{e}cnico, Universidade de Lisboa, 1049-001 Lisboa, Portugal}
\affiliation{Inductiva Research Labs, Rua da Prata 80, 1100-420 Lisboa, Portugal}
\author{Christoph Weniger}
\affiliation{GRAPPA Institute, Institute for Theoretical Physics Amsterdam and Delta
Institute for Theoretical Physics,
University of Amsterdam, Science Park 904, 1098 XH Amsterdam, The Netherlands}


\begin{abstract}
Axions can be copiously produced in localized regions of neutron star magnetospheres where the ambient plasma is unable to efficiently screen the induced electric field. As these axions stream away from the neutron star they can resonantly transition into photons, generating a large broadband contribution to the neutron star's intrinsic radio flux. In this work, we develop a comprehensive end-to-end framework to model this process from the initial production of axions to the final detection of radio photons, and derive constraints on the axion-photon coupling, $g_{a\gamma\gamma}$, using observations of 27 nearby pulsars. We study the modeling uncertainty in the sourced axion spectrum by comparing predictions from 2.5 dimensional particle-in-cell simulations with those derived using a semi-analytic model; these results show remarkable agreement, leading to constraints on the axion-photon coupling that typically differ by a factor of no more than $\sim 2$. The limits presented here are the strongest to date for axion masses $10^{-8} \eV \lesssim m_a \lesssim 10^{-5} \eV$, and crucially do not rely on the assumption that axions are dark matter.
\end{abstract}

\maketitle

\section{Introduction}
Axions are among the best-motivated candidates for physics beyond the Standard Model; these particles are a fundamental prediction of the leading solution to the strong CP problem~\cite{PQ1, PQ2, WeinbergAxion, WilczekAxion}, are an ideal candidate to explain the `missing matter' in the Universe (\ie dark matter)~\cite{PRESKILL1983127,Abbott1982,Fischler1982, collaboration2020planck}, and are expected to arise naturally in string theory from the compactification of gauge fields on topologically nontrivial manifolds~\cite{Svrcek_2006,Axiverse2010}. 

Axions generically couple to electromagnetism via the Lagrangian term $\mathcal{L} = -\frac{1}{4}\,g_{a \gamma \gamma}\,\vec{E}\,\cdot\,\vec{B}\,a$, where $a$ is the axion, $\vec{E}$ and $\vec{B}$ are the electric and magnetic fields, and $g_{a\gamma\gamma}$ is a dimensionful coupling constant. This interaction allows for axions and photons to mix in the presence of external magnetic fields, a process which is searched for 
in both laboratory experiments (\eg  axion haloscopes~\cite{Sikivie1983,DePanfilis:1987, Hagmann:1990,Hagmann:1998cb,Asztalos:2001tf, Asztalos:2009yp,Du:2018uak,Braine2020,Bradley2003,Bradley2004,Shokair2014,HAYSTAC, Zhong2018,Backes_2021,mcallister2017organ,QUAX:2020adt,Choi_2021,TheMADMAXWorkingGroup:2016hpc,Majorovits:2017ppy,MashaCurlyRobert2018}, helioscopes~\cite{CAST2009,CAST2015, Anastassopoulos2017} and light-shining-through-walls experiments~\cite{Bibber1987,Rabadan2006,Adler2008}) and in indirect astrophysical searches, such as those looking for spectral features in X-rays and gamma-rays (see \eg \cite{Wouters:2013hua,HESS:2013udx,Payez_2015,Fermi-LAT:2016nkz,Meyer:2016wrm,Marsh:2017yvc,Reynolds:2019uqt,Xiao:2020pra,Li:2020pcn,Dessert:2020lil,Meyer:2020vzy,Calore:2021hhn,Reynes:2021bpe}) and radio searches for spectral lines~\cite{Foster:2020pgt,battye2021robust,Foster:2022fxn}\footnote{Note that there also exist searches for radio lines from axion decay~\cite{Caputo:2018ljp,Caputo:2018vmy,Ghosh:2020hgd,Buen-Abad:2021qvj,Sun:2021oqp}, however the detection prospects are significantly weaker. }.

One environment in which axion-photon mixing is particularly strong is the magnetosphere of a neutron star, where the large coherent magnetic field and the ambient plasma allow for highly efficient resonant transitions\footnote{Resonant transitions occur when an axion/photon traverses a medium where the axion and photon momentum are approximately equal, \ie $k_a \simeq k_\gamma$.}. If axions contribute to the dark matter, one expects this conversion to generate narrow spectral lines that can be observed using radio telescopes~\cite{Pshirkov:2007st, Huang:2018lxq, Hook:2018iia, Safdi:2018oeu, Battye:2019aco, Leroy:2019ghm, Foster:2020pgt, Buckley2021, Witte2021, Battye2021, battye2021robust, Foster:2022fxn}; this idea has ignited numerous observational efforts~\cite{Foster:2020pgt, battye2021robust,Foster:2022fxn}, with the most recent study setting world-leading limits for axion masses near $\sim 30 \, \mu{\rm eV}$~\cite{Foster:2022fxn}. Despite the success and future promise of these spectral line searches, they are limited by the assumption that axions contribute significantly to dark matter, that the dark matter is smoothly distributed throughout the Galaxy, and that the Galactic dark matter halo is cuspy. Furthermore, such searches are confined to masses $10^{-7} \eV \lesssim m_a \lesssim 10^{-4} \eV$, as the mass must be sufficiently high to produce observable radio emission and sufficiently low so that resonances may be encountered.

\begin{figure*}
    \centering
    \includegraphics[width=.8\textwidth]{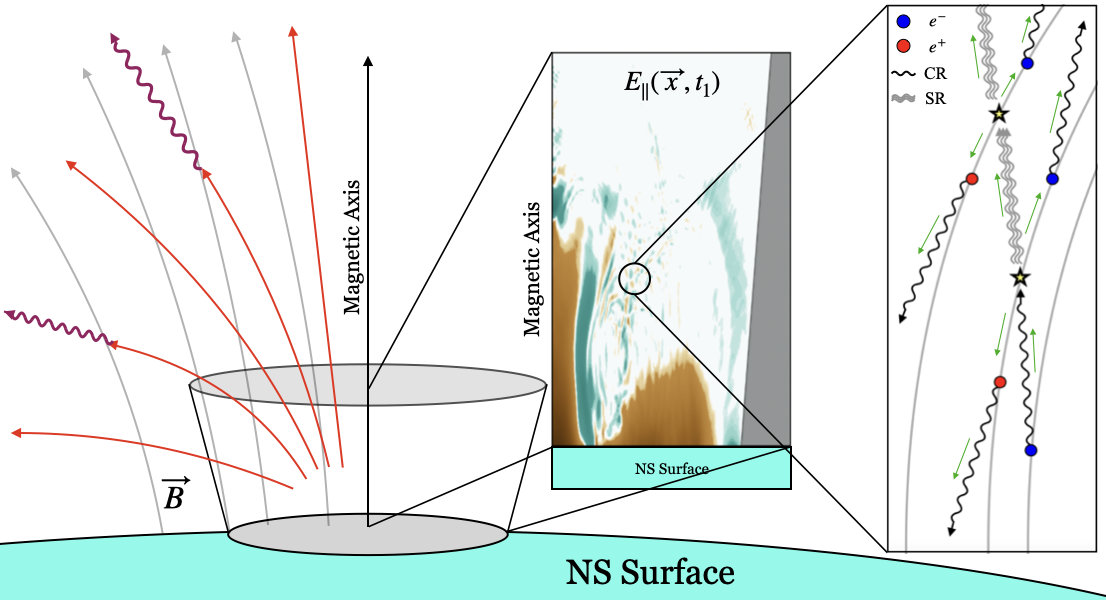}
    \caption{Schematic figure showing axion production in neutron star vacuum gaps. The vacuum gap is depicted by a truncated cone on the neutron star surface. The left inset shows a time snapshot of $\epar$ (from the simulations of \cite{Cruz2021}), with the brown/green coloring reflecting negative/positive values of $\epar$. The right inset depicts the microphysical processes responsible for the pair cascade, with green arrows indicating the direction the cascade flows with time. Axions (red) are emitted from the gap and convert to photons (purple) in the presence of the neutron star's magnetic field, $\vec{B}$ (gray).}
    \label{fig:key_fig}
\end{figure*}

Recently, Ref.~\cite{Prabhu2021} proposed an alternative way to detect axions in neutron star magnetospheres that overcomes the aforementioned challenges. The idea is based on relativistic axions sourced locally in the magnetosphere from the spacetime oscillations of $\vec{E} \cdot \vec{B}$\footnote{Axions may also be produced at the neutron star rotational frequency if there is a large-scale unscreened $\vec{E} \cdot \vec{B}$~\cite{Garbrecht:2018akc}.}. Ref.~\cite{Prabhu2021} showed that the electromagnetic fields are strong enough in the polar caps of neutron stars (situated above the magnetic poles, and spanning distances of order $r_{\rm pc} \sim \mathcal{O}(10-100)$ meters -- see Fig.~\ref{fig:key_fig}) to produce an enormous flux of axions. As these axions traverse the magnetosphere they may encounter resonances, generating a broadband radio flux in the $\MHz - \GHz$ regime. This axion-induced radio flux provides an alternative observable in the search for these evasive particles.

In this work, we construct the first pipeline to compute the intrinsic spectrum of axions produced in neutron star polar caps, their resonant conversion to photons, and the nonlinear evolution of these radio photons as they escape the magnetosphere. Our analysis uses both state-of-the-art numerical simulations as well as a newly developed semi-analytic model to predict the axion production rate; the overall agreement of these approaches illustrates that our procedure is not strongly sensitive to reasonable modeling uncertainties in the gap dynamics. We use this pipeline to constrain the axion-photon coupling by comparing the predicted radio flux with measurements of 27 nearby pulsars. The constraints derived here are the strongest to date for axion masses spanning $10^{-8} \eV \lesssim m_a \lesssim 10^{-5} \eV$.

\section{Axion Production from Vacuum Gap Discharges}
The $e^\pm$ pair plasma populating the magnetospheres of neutron stars is expected to efficiently screen the component of the background electric field along the magnetic field lines ($\epar$), except in small localized regions referred to as vacuum gaps which are responsible for particle acceleration and production of the pair plasma itself~\cite{GoldreichJulian1969,Romani1992}. 

Vacuum gaps are expected to arise in a variety of locations, including in the polar caps~\cite{RudermanSutherland1975}, the slot gap (located near the neutron star along the last closed field lines)~\cite{Arons1983}, and the outer gap (located near the light cylinder along the last closed field lines)~\cite{Cheng1986}. 
Recent progress in global particle-in-cell (PIC) simulations of the magnetosphere has shown that $e^\pm$ pairs can be produced in the current sheet near the light cylinder, and efficiently screen gaps situated along the last closed field lines~\cite{Brambilla2018,Hu2021}.
We therefore choose to focus on the dynamics of the vacuum gaps in the polar caps, which also have the highest local values of $\vec{E}\cdot\vec{B}$.

Recently, local PIC simulations of the neutron star polar-cap pair cascade have been performed~\cite{TimokhinArons2013,Philippov2020,Cruz2021}. These simulations show that the discharge is an oscillatory process: $e^\pm$ pairs accelerated in the gap produce gamma-rays that convert to more $e^\pm$ pairs, which proceed to screen the gap, shutting down pair production~\cite{Levinson2005}. It was proposed that this process can produce coherent electromagnetic radiation, potentially answering the long-standing puzzle of the origin of pulsar radio emission~\cite{Philippov2020}.

The oscillatory pair discharge process will in general lead to inductively driven oscillations of $\epar$ at a frequency set by the local plasma frequency, which is expected to evolve from the Goldreich-Julian (GJ) oscillation frequency $\omega_{\rm GJ} = \sqrt{4 \pi \alpha \, n_{\rm GJ} / m_e}$\footnote{We work in units where $c = \hbar = 1$.} to a value $\mathcal{O}(10 - 100)$ times greater as the density of the pair plasma increases~\cite{Levinson2005, TimokhinHarding2015, Fabio2021, Tolman:2022unu}. Here we have have introduced the GJ charge density, given by $n_{\rm GJ} \equiv 2 \vec{B_0} \cdot \vec{\Omega}_{\rm NS} / e$, with $B_0$ and $\Omega_{\rm NS}$ being the surface magnetic field strength and rotational frequency of the neutron star, $e$ the elementary charge, $m_e$ the electron mass, and $\alpha$ the fine structure constant. The presence of an oscillating $\epar$ directly enters as a source term in the axion's equation of motion,
\begin{equation}\label{eqn:KG} 
    (\Box + m_a^2) a(x) =  - \gagg (\vec{E} \cdot \vec{B})(x) \, .
\end{equation}
The differential production rate of axions per momenta $\vec{k}$ can subsequently be expressed as~\cite{peskin2018introduction}
\begin{equation}\label{eqn:diffrate}
    \frac{d \dot{N}}{d^3 k}  = \frac{\left| \tilde{\jmath}(\vec{k}) \right|^2}{2 (2\pi)^3 \omega(\vec{k}) T} \, ,
\end{equation}
where $\omega(\vec{k})$ is the axion energy, $T$ is the quasiperiodic timescale of the gap collapse, and $\tilde{\jmath}(\vec{k})$ is the Fourier transform (FT) of the source term,
\begin{equation}\label{eq:fftj}
    \tilde{\jmath}(\vec k) = \displaystyle - \gagg \int d^4 x \, e^{i k \cdot x} (\vec{E} \cdot \vec{B})(x) \,.
\end{equation}
The rate and spectrum of axions produced during the collapse of the vacuum gap is fully determined by the spacetime evolution of $\epar$. The condition that axions be on shell inherently connects the spatial and temporal evolution of $\epar$, meaning a given $k$-mode can only be produced if the spatial component of the FT contains support on scales $\sim k^{-1}$ \emph{and} the temporal component contains support on scales $\sim (k^2 + m_a^2)^{-1/2}$.

In order to provide a quantitative understanding of axion production, we briefly estimate the production rate for a pulsar with $B_0 = 10^{12} \, \rm G$ and $\Omega_{\rm NS} = 2 \pi \, \rm Hz$. The polar-cap radius for such a pulsar is $r_{\rm pc} \sim 150 \, \rm m$, and the maximum value of the unscreened electric field is roughly $\epar \sim 6\times10^{-6} \, B_0$, see the Supplementary Material (SM). We expect production to be most efficient when $\epar$ is largest, which occurs prior to the screening phase when the characteristic scale is $k_c \sim 2\pi / r_{\rm pc} \sim 10^{-8} \, {\rm eV}$. Neglecting the phase in Eq.~\eqref{eq:fftj}, and taking $d^3 k / \omega \sim k_c^2$, $T \sim 10^{-7} \, \rm s$, and $g_{a\gamma\gamma} \sim 10^{-11} \, {\rm GeV}^{-1}$, we find $\dot{N} \sim \mathcal{O}(10^{50})$ axions per second, which is comparable to the values obtained from more careful calculations in the SM.

\section{Modeling Gap Collapse}
We adopt two approaches to estimate the axion spectrum produced from the gap dynamics, one based on a semi-analytic model and the other on a numerical simulation.

\subsection{Semi-analytic Model} 
We begin by highlighting the key physical features entering our 2+1D semi-analytic model; a more detailed description of each step is deferred to the SM.

We model two stages of gap evolution: pair cascade and gap collapse. The pair cascade phase begins with an initially unscreened $\epar$ and low charge density. As the pair cascade progresses, the number density increases exponentially until it reaches $n_\text{GJ}$. At this point the gap locally collapses, marking the end of the pair cascade. 

A single seed particle (electron) is assumed to initiate the pair cascade. Because of the presence of $\epar$, the electron is accelerated to the radiation-reaction-limited Lorentz factor, $\gammamax$, in time $\tacc$. Since particles move along curved field lines, they then emit curvature radiation (CR) photons with characteristic energy $ \varepsilon_\text{CR} \propto \gammamax^3/\rho_c$, where $\rho_c$ is the field line curvature \cite{Jackson, TimokhinHarding2015}. CR photons may be absorbed by the magnetic field to produce new pairs and synchrotron photons. Newly produced electrons (positrons) are accelerated away from (towards) the neutron star surface, producing their own CR photons, which may again produce new pairs. Synchrotron photons can also produce new pairs if their mean free path is less than the size of the gap (see Fig.~\ref{fig:key_fig}). 

In our model, we compute the creation positions, times, and energies of all photons and pairs produced by the single seed particle. Our model iterates over `generations’ of particles, where the first generation is the seed particle, and generation $n$ particles are sourced by generation $n-1$. We run five generations of the cascade for different seed particle locations within the gap, identifying points where the plasma density locally reaches $n_\text{GJ}$ as the starting points for gap collapse -- these points, which we refer to as `pair production seeds', define the initial conditions of the semi-analytic model (see SM).  

The initial stages of the gap collapse have been studied using analytic toy models and numerical simulations; in both cases, one expects the initial burst of particle production to induce exponentially damped oscillations in $\epar$. These oscillations initially occur locally, and subsequently propagate outwards along the magnetic axis~\cite{Levinson:2005fa,Timokhin:2010fe,Levinson:2018arx,Chen:2019osy,Philippov2020,Kisaka:2020lfl,Cruz:2020vfm,Cruz2021,Tolman:2022unu,Cruz:2022zyp}. The frequency of the damped oscillations is set by the local plasma frequency (which itself is set by the charge density and typical Lorentz factor), and will evolve from $\omega_{\rm GJ}$ (\ie the minimum value necessary to collapse the gap) to a value $10 - 100$ times greater as more pairs are produced. The characteristic screening timescale $t_c$ is set by the time required for pair production processes to yield a GJ charge density, which is roughly equivalent to the time required to accelerate charges to $\gamma_{\rm max}$~\cite{Tolman:2022unu}; for realistic pulsars this timescale lies between $10^{-9} - 10^{-6} \, \rm s$.

In order to capture the general features of the gap collapse process, we model the unscreened electric field $\epar$ with a static profile, and describe the screening of $\epar$ as the combination of outward propagating 2D plane waves. Each plane wave is exponentially damped on a timescale $t_c$, and has a time-evolving oscillation frequency growing from an initial value $\omega_{\rm GJ}$. In total, we source four propagating plane waves\footnote{This number must be sufficiently large to cover the gap, but cannot be too large as interference effects are not properly accounted for in this treatment. In the SM we show that taking 3 or 5 plane waves leads to a negligible difference.} evenly spaced across the width of the gap, and with initial conditions determined by the one-dimensional pair-production process discussed above. 

\subsection{PIC Simulation}
Our second model used to compute the axion spectrum relies on a 2.5 dimensional\footnote{2.5D means using azimuthal symmetry to reduce the problem to 2D, but still evolving all 3 components of vector quantities.} PIC simulation developed in~\cite{Fabio2021}. Working in axisymmetric cylindrical coordinates $(r,z)$, the authors impose a dipolar magnetic field. A rotating disk of radius $r_{\rm pc}$ is established at the stellar surface to produce a potential drop in the open field line zone of the neutron star. Outside $r = r_{\rm pc}$, $\epar$ is forced to zero to model the plasma-filled closed field line zone. Particles are extracted from the surface at a rate that depends on the local value of $\epar$; these particles are accelerated to the radiation-reaction limit and emit gamma-rays through synchrocurvature radiation. Those gamma-rays subsequently produce $e^{\pm}$ pairs in the ultrastrong magnetic field through one-photon magnetic pair production~\cite{Erber1966}. This process is modeled using the state-of-the-art QED module in the PIC code OSIRIS~\cite{fonseca2002osiris}. Videos showing the dynamical screening of $\epar$ are available \href{https://www.youtube.com/watch?v=wHSw5kp7Ik4}{online}; a snapshot of the simulation is shown in the left inset of Fig.~\ref{fig:key_fig}.

Performing simulations from first principles of the gap collapse process is extremely challenging due to the large separation of scales between the size of the polar cap and the kinetic scale of the plasma (typically differing by $4 - 5$ orders of magnitude). The simulations performed in~\cite{Fabio2021} overcame this difficulty by re-scaling the quantum parameters $\chi_{\pm,\gamma} = \sqrt{\left(p_\mu F^{\mu\nu}\right)^2}/(B_c m_e)$  for both photon emission and pair production, multiplying them by a numerical constant; here, $p_\mu$ is the 4-momentum of the corresponding $e^\pm$ or $\gamma$, and $B_c \simeq 4.4 \times 10^{13} \, \rm G$ is the Schwinger field strength. This re-scaling effectively allows pair production to occur at a much lower voltage drop. However, it also significantly reduces the inherent scale separation in the problem: the kinetic plasma length scale becomes only a few hundredths of the polar-cap size. As a result of this compression of scales, the FT necessarily compresses the power to an artificially narrower range of $k$-modes (the largest scales are expected to be unaffected, however power in small scales, \ie large $k$-modes, will have been shifted to intermediate scales). We describe in Section~\ref{app:model} of the SM a procedure for re-scaling  the power of the FT, along with additional details on the PIC simulation.

\section{The Radio Flux}
Once the initial axion spectrum has been determined, we employ an updated version of the ray-tracing algorithm developed in~\cite{Witte2021} to compute the radio spectrum. We describe the general features of this procedure below, and defer further details to the SM. 

We begin by calculating the rate of axion production for momenta spanning from the escape momentum to $\sim \mathcal{O}(10 \, {\rm GHz})$ (the phenomenology of gravitationally bound axions differs markedly, and thus we leave a detailed study of bound states to a companion paper~\cite{BoundStates}). For each momentum state the corresponding axion trajectory is propagated to the light cylinder, and all resonances encountered during propagation are identified. Resonances occur when~\cite{Witte2021, millar2021axionphotonUPDATED}
\begin{equation}
    \omega_p^2 \simeq \frac{m_a^2 \omega^2}{m_a^2 \cos^2\theta + \omega^2 \sin^2\theta} \, ,
\end{equation}
where $\theta$ is the angle between the axion momentum and the magnetic field.

At every level-crossing we compute the conversion probability, $P_{a\rightarrow\gamma}$, and axion survival probability\footnote{We do not compute the axion production probability for subsequent resonances encountered by sourced photons, as this quickly becomes computationally prohibitive. For the pulsars and couplings studied here, we do not expect this secondary contribution to be significant.}, $P_{a\rightarrow a}  = 1 - P_{a\rightarrow \gamma}$, and subsequently weight each photon sourced at crossing $i$ by $P_{a\rightarrow\gamma}^{(i)} =P_{a\rightarrow\gamma} \times \Pi_{j=1}^{i-1} P_{a\rightarrow a}^{(j)}$. The conversion probability at an individual level crossing is given by the Landau-Zener formula~\cite{Battye:2019aco}
\begin{equation}
P_{a\rightarrow \gamma} = 1 - e^{-\Gamma} \, ,
\end{equation}
with
\begin{equation}
\Gamma = \frac{\pi}{2} \Big(1 + \frac{\omega_p^4 \Delta^2 \cos^2\theta}{\omega^4}\Big) \Big(\frac{\omega g_{a\gamma\gamma} B \Delta}{k_a}\Big)^2 \frac{1}{\lvert \partial_s k_\gamma \rvert} \, .
\end{equation}
Here, we have introduced $\Delta \equiv \sin\theta / (1 - \omega_p^2 \cos^2\theta / \omega^2)$
and defined~\cite{millar2021axionphotonUPDATED} 
\begin{equation}
\partial_s \equiv \partial_{\hat{k}_{\parallel}} - (\omega_p^2 \Delta \cos\theta / \omega^2) \partial_{\hat{k}_{\perp}} \, ,
\end{equation}
with $\hat{k}_{\parallel}$ ($\hat{k}_{\perp}$) representing the parallel (perpendicular) direction to the axion momentum. Photons are propagated to the light cylinder using the dispersion relation for a Langmuir-O mode (see~\cite{Witte2021}). The observed spectrum is obtained by summing over the final distribution of binned and weighted photons -- example spectra are included in the SM.

\begin{figure}[t!]
    \includegraphics[width=.48\textwidth]{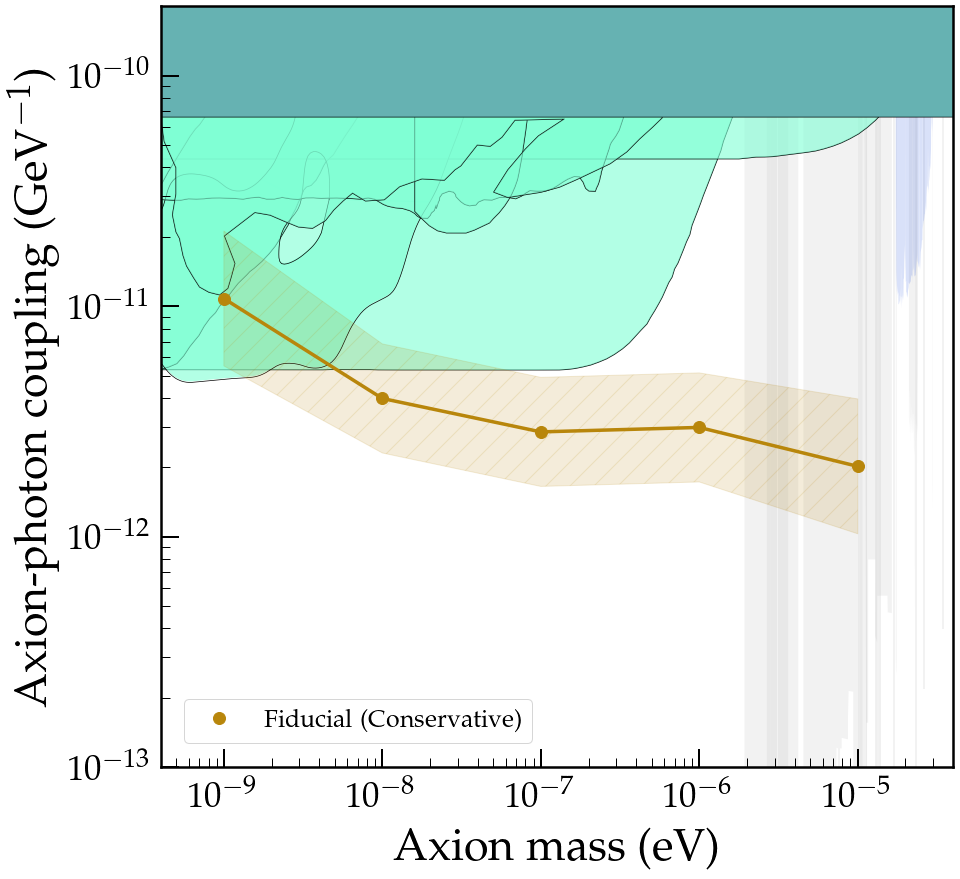}
    \caption{Upper limits on the axion-photon coupling derived in this work using a combination of a 2.5D PIC simulation and the semi-analytic model. The band reflects a conservative estimate of the modeling uncertainties (see SM). We compare to existing constraints from neutron stars~\cite{Foster:2022fxn} (blue), haloscopes~\cite{Sikivie1983,DePanfilis:1987,Hagmann:1990,Hagmann:1998cb,Asztalos:2001tf,Asztalos:2009yp,Du:2018uak,Braine2020,Bradley2003,Bradley2004,Shokair2014,HAYSTAC,Zhong2018,Backes_2021,mcallister2017organ,QUAX:2020adt,Choi_2021,Alvarez_Melcon_2021} (gray), helioscopes~\cite{Anastassopoulos2017} (teal), and astrophysics~\cite{Wouters:2013hua,HESS:2013udx,Payez_2015,Fermi-LAT:2016nkz,Meyer:2016wrm,Marsh:2017yvc,Reynolds:2019uqt,Xiao:2020pra,Li:2020pcn,Dessert:2020lil,Dessert1_2022,Dessert2_2022} (light green). The former two have reduced opacity to highlight that they rely on axions being dark matter. }
    \label{fig:Limits}
\end{figure}

\section{Results and discussion}
The radio emission mechanism of active pulsars is not well understood, making it difficult to identify signatures arising from this process. Nevertheless, one can constrain the existence of axions without knowing the intrinsic pulsar flux by requiring that their contribution not exceed the observed flux. In this section we derive limits on $g_{a\gamma\gamma}$ using observations of 27 nearby pulsars; our sample includes representative nearby pulsars, whose surface magnetic field and rotational period roughly span $10^{12} \, {\rm G} \lesssim B_0 \lesssim 10^{13} \, \rm G$ and $10^{-1} \, {\rm s} \lesssim P_{\rm NS} \lesssim 2 \, \rm s$. For computational ease, we choose to focus on pulsars whose radio emission geometry is constrained by observations, thus evading the need to marginalize over the misalignment and viewing angles. Details of all pulsars used in this analysis are presented in the SM.

We present our fiducial $95\%$ confidence level upper limits in Fig.~\ref{fig:Limits}, which are obtained by computing the constraints for both models (a comparison between models is left to the SM) and taking the weaker limit at each mass. The bands around the fiducial limit represent a conservative estimate of the systematic uncertainties (see SM). For comparison, we plot constraints from radio line searches using neutron stars (blue), axion haloscopes (gray), the CAST experiment (teal) and X-ray and gamma-ray telescopes (light green). The limits derived in this work significantly improve upon existing bounds, and unlike axion haloscope experiments (and radio line searches), do not assume axions contribute to the dark matter. In addition, since the radio flux scales $\propto g_{a\gamma\gamma}^4$, the constraint is largely insensitive to minor mismodeling errors. The mass range covered by our constraints is limited by the frequency of radio observations (higher frequencies could probe higher masses), and the computational expense (computing time increases at both lower and higher masses).

In the SM we show that the derived bound is controlled by observations of a few strong pulsars, with high frequency observations providing the most constraining power. A comprehensive analysis of all pulsars in the ATNF catalog, as well as more dedicated pulsar observations at high frequencies, could significantly improve upon these results; we reserve this broader analysis for future work.

\acknowledgments{}
We would like to thank Georg Raffelt, Andrea Caputo, Ben Safdi, and Anatoly Spitkovsky for useful discussions. The authors would also like to thank Andrea Caputo and Jamie McDonald for their useful comments on the draft. DN, SJW and CW are supported by the European Research Council (ERC) under the European
Union's Horizon 2020 research and innovation programme (Grant Agreement No. 864035 - Undark). AP acknowledges support from the National Science Foundation under Grant No. PHYS2014215, and from the Gordon and Betty Moore Foundation Grant No. GBMF7946. AC acknowledges support from Fermi Guest Investigation Grant No. 80NSSC21K2027 and NSF Grant No. DMS-2235457. FC acknowledges support from the European Research Council (InPairs ERC-2015-AdG 695088), FCT (PD/BD/114307/2016 and APPLAuSE PD/00505/2012), and PRACE for awarding access to MareNostrum (Barcelona Supercomputing Center, Spain), where the PIC simulations used in this work were performed. We acknowledge the use of~\cite{AxionLimits} in creating the figures containing axion constraints.

\bibliography{gap_axion}

\clearpage

\onecolumngrid

\vspace{1cm}
 
\begin{center}
  \textbf{\large Supplementary Material for Novel Constraints on Axions Produced in Pulsar Polar-Cap Cascades}\\[.2cm]
  \vspace{0.05in}
  {Dion Noordhuis, Anirudh Prabhu, Samuel J. Witte, Alexander Y. Chen, F\'{a}bio Cruz, and Christoph Weniger}
\end{center}

\setcounter{equation}{0}
\setcounter{figure}{0}
\setcounter{table}{0}
\setcounter{section}{0}
\setcounter{page}{1}
\thispagestyle{empty}
\makeatletter
\renewcommand{\theequation}{S\arabic{equation}}
\renewcommand{\thefigure}{S\arabic{figure}}
\renewcommand{\thetable}{S\arabic{table}}
\def\set@footnotewidth{\onecolumngrid}

This Supplementary Material provides additional details and results for the analyses discussed in the main Letter.

\section{Modeling Vacuum Gap Collapse}\label{app:model}
In this section we illustrate specifics concerning the computation of the initial axion spectra. We begin by discussing in detail the relevant PIC simulation developed in~\cite{Fabio2021}, and then describe the ingredients used in the semi-analytic model. 

\subsection{PIC Simulation}
We use the results from one of the 2.5 dimensional QED particle-in-cell simulations developed in~\cite{Fabio2021} and performed using the code OSIRIS~\cite{fonseca2002osiris,fonseca2008one}. Here, we outline the details of this simulation, introduce the re-scaling methodology, and explain how we extrapolate the simulation results back to more realistic regimes.

The simulation is performed in axisymmetric cylindrical coordinates $(r,z)$, where $r = 0$ is the rotation axis and $z=0$ is the stellar surface (ignoring the curvature of the surface) modelled as a conductor. The stellar magnetic field is taken to be a dipole with magnetic axis aligned with the rotation axis, and magnitude given by $B(r=0, z=0) = 0.1 \times B_c$ with $B_c$ the critical magnetic field strength (conventionally, $B_c \simeq 4.4\times 10^{13} \, \mathrm{G}$, however in the simulation $B_c$ is reduced to $10^7 \, \textrm{G}$ to make the computations feasible). To mimic the potential drop induced by the rotation of the neutron star, we rotate a disk of radius $r_{\rm pc}$ on the stellar surface; $r_{\rm pc}$ is known as the polar-cap radius and given by
\begin{equation}
    r_{\rm pc} \simeq R_{\rm NS} \, \sqrt{R_{\rm NS} \Omega_{\rm NS}} \, ,
\end{equation} 
where $R_{\rm NS}$ is the neutron star radius and $\Omega_{\rm NS}$ its rotational frequency. The polar-cap radius is chosen to be $0.1R_{\rm NS}$ in the simulation, which corresponds to $\Omega_{\rm NS}$ of a few hundred Hz. Rotation of the disk is achieved by imposing a boundary condition on the radial electric field $E_r(r, z=0) = E_0 \times (r/r_{\rm pc}) \times g(r / r_{\rm pc})$, with $E_0 = \Omega_{\rm NS} B_0 R_{\rm NS}$ and $g(x) = 0.5 \times (1 - \tanh( (x-1)/ 0.2))$. The latter function has been chosen to allow for a smooth transition from the open to the closed field line zone. The upper boundary of the computational domain is taken to be open for both particles and fields~\cite{fonseca2002osiris, Cruz2021}. The plasma-filled closed field line zone is modelled as a perfect conductor where $E_\parallel = 0$.

The simulation is performed on a grid that spans $L_r \times L_z = 1.5 \, r_{\rm pc} \times 2.5 \, r_{\rm pc}$, with resolution $6000 \times 10000$. This resolution, however, does not allow us to resolve the realistic scale separation between the polar-cap size $r_{\rm pc}$ and the plasma skin depth $\lambda_p$, especially when pair production creates a large number of $e^\pm$ particles. In order to resolve the physics correctly, the plasma skin depth corresponding to the GJ charge density, $\lambda_\mathrm{GJ} = \omega_{\rm GJ}^{-1}$, is increased to a value of $\sim 5\times 10^{-3} \, r_{\rm pc}$, roughly $\mathcal{O}(100)$ times larger than expected in realistic neutron stars. This re-scaling decreases the overall potential drop induced in the open field line bundle. Nevertheless, in order to model the pair production physics the electrons must still be allowed to create pairs at the lower energies. This is achieved by also scaling the quantum parameters $\chi_\pm$ and $\chi_\gamma$, which govern the emission of gamma-ray photons and the rate of conversion to $e^\pm$ pairs. These quantum parameters are defined by $\chi_{\pm,\gamma} = \sqrt{\left(p_\mu F^{\mu\nu}\right)^2}/(B_c m_e)$ -- here $p_\mu$ is the 4-momentum of the corresponding particle, $F^{\mu\nu}$ the electromagnetic field tensor, and $m_e$ the electron mass -- and are multiplied by a factor of $\zeta_{\pm,\gamma}=10^3$ in the simulation to enhance pair production at lower energies.

The simulation proceeds as follows. At each time step in the simulation charges are injected at the surface with an effective charge density proportional to the local parallel electric field, $n_{\rm inj} = \kappa (\epar /e r_{\rm pc})$, with $\kappa = 0.1$ and $e$ the unit of elementary charge. As the particles are accelerated from the surface, their quantum parameters $\chi_\pm$ are individually calculated, multiplied by the scaling factor $\zeta_\pm$, and then used to emit synchrocurvature photons with energies sampled via Monte-Carlo methods. Upon emission of a synchrocurvature photon, the energy of the photon is subtracted from that of the emitting particle. This is a discrete implementation of the `radiation-reaction force' (see the following subsection for additional discussion). The emitted photons are also propagated in the simulation, their quantum parameters $\chi_\gamma$ are computed, multiplied by the scaling factor $\zeta_\gamma$, and then used to compute the quantum Breit-Wheeler pair production cross-section. The photon is afterwards conditionally converted to an $e^\pm$ pair based on the resulting pair production probability. The initial conditions of the simulation are set such that there are no particles along the open field lines. The conducting disk is gradually spun up, releasing particles which subsequently initiate quasiperiodic particle cascades. When performing the FT, we remove the initial part of the simulation, and focus only on the quasiperiodic solution achieved in the later stages. 

The re-scaling of fundamental parameters mentioned above introduces a complication -- the plasma scale is artificially inflated with respect to the size of the system (instead of being separated by $\sim 4 - 5$ orders of magnitude, they are only separated by $\sim 3$ orders of magnitude). When calculating the FT, this re-scaling shifts power from $k$-modes near $k \sim \lambda_{\rm GJ}^{-1}$ to a value $\mathcal{O}(100)$ times smaller. In order to correct for this effect we compute the FT $\tilde{\jmath}$ at a log-spaced series of momentum modes $(\tilde{k}_{\rm min}, \cdots, \tilde{k}_{\rm max})$, and subsequently re-stretch the momentum vector $k_i = (\xi)^{i/n} \, \tilde{k}_i$ before computing the differential rate in Eq.~\eqref{eqn:diffrate}. Here $k_{\rm min}$ and $k_{\rm max}$ are determined by the largest and smallest scales resolvable by the semi-analytic model, and $n$ is the total length of the series. The extrapolation factor $\xi$ is found by computing the ratio between the GJ scale predicted using the neutron star parameters and the value inferred from the ratio of $\lambda_{\rm GJ} / r_{\rm pc}$ used in the simulation. 

Finally, note that, due to the parameter re-scalings highlighted above, one must re-scale the value of $E_{||}/ B_0$ from the simulation to each pulsar under consideration. In order to do this, we use the fact that $E_{||}/B_0 \propto \Omega_{\rm NS} r_{\rm pc} \propto \Omega_{\rm NS}^{3/2}$. In the simulation we analyzed, the parallel electric field, $E_{||}^s$, corresponds to an effective rotational frequency of $\Omega_{\rm NS}^{s,{\rm eff}}\sim 200 \, {\rm Hz}$. Keeping the above in mind, we use the following re-scaling relation to determine the parallel electric field, $E_{||}^p$, for any real pulsar
\begin{equation}
    E_{||}^p = \left(\frac{\Omega_{\rm NS}^p}{\Omega_{\rm NS}^{s, {\rm eff}}} \right)^{3/2} \frac{B_0^p}{B_0^s} E_{||}^s \, .
\end{equation}

\subsection{Semi-analytic Model}
There are two primary reasons to jointly develop a semi-analytic model. Firstly, the simulation discussed above was developed for a particular neutron star. We have applied it to the more generic population by re-normalizing the polar-cap radius and the maximum amplitude of $\epar$ to their theoretical values, but it is not guaranteed that this re-normalization is sufficient to capture all relevant features. The second concern is that the re-scaling procedure introduced in the simulation to reduce scale separation may generate unexpected features in the FT. In order to address both of these concerns, we have built a semi-analytic model that can: $(i)$ be applied to the broader class of pulsars and $(ii)$ be run on a higher resolution grid, significantly reducing (although not entirely eliminating) the need for post-analysis stretching of the FT. We describe the details of this semi-analytic model below.

{\bf Pair cascade: } An initially unscreened vacuum gap becomes unstable to runaway pair production. Due to variation in the magnetic field strength and line curvature across the gap pair production occurs non-uniformly. Once the local charge density becomes comparable to the GJ value dynamical screening of $\epar$ takes place, leading to exponential damping of the parallel electric field. The dynamics of $\epar$ in 1+1 dimensions are described in~\cite{Tolman:2022unu,Cruz:2022zyp} and summarized in the following section. In order to extend these results to 2+1 dimensions, we construct a 2+1 semi-analytic model of pair cascades in the gap. The model provides the initial conditions for sourcing the aforementioned damped oscillations. 

We use axisymmetric cylindrical coordinates with the lower boundary being a disk of radius $\rpc$ rotating at frequency $\Omega_{\rm NS}$. The initial parallel electric field is constant throughout the gap with magnitude $\epar = 2 \Omega_{\rm NS} B_0 \rpc$, where $B_0$ is the surface magnetic field. The domain of the gap is also permeated with a constant magnetic field, $B_0 \hat{z}$\footnote{In reality, the magnetic field strength varies across the gap but the variation is of the order $\rpc/R_\text{NS} \ll 1$. The dominant contribution to the non-uniformity of pair production is the varying field line curvature.}. We further approximate the pair cascade initiated by a single seed particle to develop along a single magnetic field line, labeled by its radial coordinate $s$. The characteristic radius of a flux tube containing the particles produced in a cascade is of the order $\psi \sim 1/\varepsilon_\gamma$, where $\psi$ is the angle between the photon momentum and the local magnetic field (\ie pitch angle) and $\epsilon_\gamma$ is the photon energy (in units of $m_e$). Low energy photons that can have larger pitch angles generally have mean free paths greater than the size of the gap, preventing them from contributing to the cascade. 

\begin{figure*}[t]
    \centering
    \includegraphics[scale=0.55]{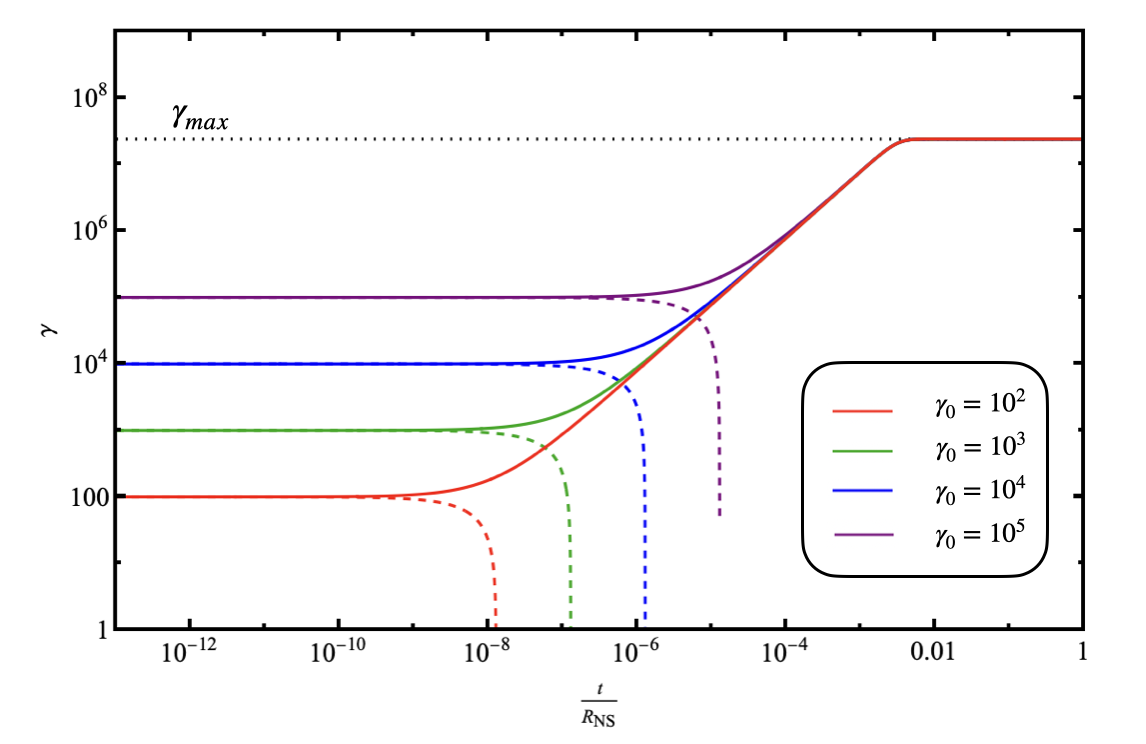}
    \caption{ Electron (solid) and positron (dashed) energies as a function of distance propagated along a curved magnetic field line with $E_0 = \Omega_{\rm NS} B_0 R_{\rm NS}$. The neutron star parameters are taken to be $b \equiv B/B_c = 0.1$, $\Omega_{\rm NS} = 2 \pi \, \rm Hz$, and $\rho_c = 100$ km. }
    \label{fig:ParticleAcceleration}
\end{figure*}

We inject macroparticles of charge $\kappa n_\text{GJ}$ at the neutron star surface, where $\kappa \ll 1$. Positrons that are accelerated away from the surface reach a maximum Lorentz factor determined by the balance between acceleration and radiative losses to curvature radiation according to the equation
\begin{align} \label{eqn:gammaevolution}
    m_e \dot{\gamma}(t) = \pm e E_0 - \frac{2 e^2}{3 \rho_c(s)^2} \gamma^4(t),
\end{align}
where $+$ ($-$) applies to positrons (electrons), $\rho_c(s) = (4/3) R_\text{NS}^2/s$ is the magnetic dipole curvature at radius $s$, and $\gamma(t)$ is the particle Lorentz factor. The first term on the RHS corresponds to acceleration by the electric field and the second term to radiative losses due to emission of curvature photons. From Eq.~\eqref{eqn:gammaevolution}, the radiation-reaction-limited Lorentz factor is $\gamma_\text{max} = \left(3 E_0 \rho_c(s)^2/2 e \right)^{1/4}$. A key quantity is the amount of time taken for a particle with initial energy $\gamma_0$ (in units of the electron mass) to reach energy $\gammamax$. This acceleration time is largely independent of the initial Lorentz factor (see Fig.~\ref{fig:ParticleAcceleration}). We define it as the amount of time it takes a particle to reach $\gamma = (1 - \epsilon) \gammamax$, which can be determined by integrating Eq.~\eqref{eqn:gammaevolution} to yield 
\begin{align}
\tacc = \frac{1}{4}\left(\frac{3 m_e^4 \rho_c(s)^2}{2 e^5 E_0^3}\right)^{1/4} \left(\pi + \ln\left(\frac{2}{\epsilon}\right) \right) \, , \label{eqn:tacc}
\end{align}
where $\epsilon \ll 1$. Going forward we set $\epsilon = 0.1$. Once a particle reaches $\gammamax$, it emits CR photons with characteristic energy $\varepsilon_\text{CR} = 3\gammamax^3/2\rho_c$ and initial pitch angle $\psi_0 = 0$. As the photon propagates its pitch angle, and hence opacity, increases. The condition for absorption by the magnetic field is~\cite{TimokhinHarding2015}
\begin{align}
    \tau(\varepsilon_\gamma, \ell_a) \equiv \displaystyle\int_0^{\ell_a} \alpha_B(\varepsilon_\gamma, \psi(x)) dx = 1 \, ,
\end{align}
where $\ell_a$ defines the mean free path and $\alpha_B$ is the opacity for pair production by a single photon in a strong magnetic field. Once the CR photon is absorbed it produces an $e^\pm$ pair, each particle having energy $\varepsilon_\gamma/2$. Since $\psi_a \equiv \psi(\ell_a)$ is non-zero at absorption the pair will be produced at a high Landau level and quickly decay to the ground state, emitting a number of synchrotron photons in the process. The final momentum (and energy assuming they are ultra-relativistic) of each fermion is given by the component of the parent CR photon energy parallel to the background magnetic field prior to absorption
\begin{align}
    \varepsilon_{\pm, f} = \frac{\varepsilon_\gamma}{2} \left(1 + \left( \frac{\chi_{\pm, a}}{b} \right)^2 \right)^{-1/2} \, ,
\end{align}
with $\chi_{\pm, a} \equiv \chi_\pm(\ell_a)$ and $b \equiv B/B_c$. We assume the remaining energy is split equally among synchrotron photons of characteristic energy $\varepsilon_\text{syn} = 3 b \psi_a \varepsilon_\gamma / 8$. Once produced, the mean free path of synchrotron photons is calculated in the same way as for the CR photons, with the exception that they inherit the pitch angle of the parent CR photon. The secondary electron (positron) is emitted upward (downward) along the magnetic field line and reaches $\gammamax$ in time $\tacc$\footnote{The positron will take slightly longer to reach $\gammamax$ since it needs to turn around and re-accelerate in the opposite direction. For the secondary particles produced the turn-around time is small compared to $\tacc$ (see dashed lines in Fig.~\ref{fig:ParticleAcceleration}), so we neglect this complication.}. The secondary electrons (positrons) can now emit upward (downward) propagating CR photons and the process continues. 

\begin{figure}
    \centering
    \includegraphics[width = 0.6\textwidth]{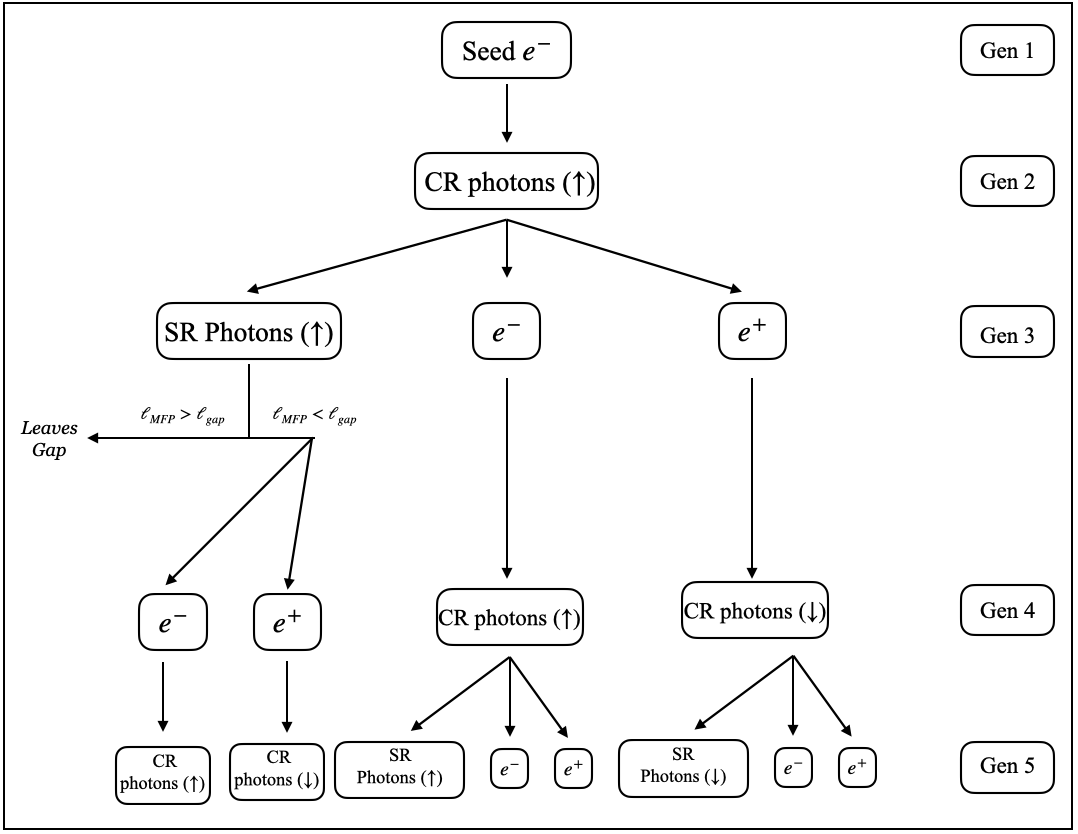}
    \caption{Schematic depiction of the generations of particle creation in our model. Photons are labeled as either upward propagating $(\uparrow)$ or downward propagating $(\downarrow)$. Each electron/positron produces many CR photons (see Eq.~\eqref{eqn:ncr} for an estimate). Each CR photon, when absorbed, produces a single electron/positron pair and several synchrotron photons. }
    \label{fig:flowchart}
\end{figure}

We study the above model using an iterative algorithm that works as follows. We define four categories of particles: electrons, positrons, upward-propagating ($\gamma^\uparrow$), and downward propagating ($\gamma^\downarrow$) photons. For each $e^\pm$ we keep track of its creation position (\zpmcreate) and time (\tpmcreate) and for each photon its creation position (\zgcreate), time (\tgcreate), energy ($\varepsilon_\gamma$), and initial pitch angle ($\psi_0$). We assume that particles reach $\gammamax$ at time $t_\text{create}+\tacc$, where $\tacc$ is given in Eq.~\eqref{eqn:tacc}, at which point acceleration is balanced by emission of CR. We assume CR photons are emitted at regular intervals $\delta t_\text{CR} = \varepsilon_\text{CR}/P_\text{CR}$, with $P_\text{CR} = 2 e^2 \gammamax^4 / 3 \rho_c^2$ being the CR power. Thus for each electron (positron)  in generation $n$ we create new upward (downward) propagating CR photons according to
\begin{align}
    t^{\pm, (n+1)}_\text{create} &= t^{\pm, (n)}_\text{create} + \tacc + i \ \delta t_\text{CR} \, , \\
    z^{\pm, (n+1)}_\text{create} &= z^{\pm, (n)}_\text{create} \pm \tacc \pm i \ \delta t_\text{CR} \, .
\end{align}
The characteristic number of CR photons emitted by each electron/positron is given by the transit time across the gap divided by the time interval between emission of CR photons,
\begin{align}\label{eqn:ncr}
    N_{\rm CR} \sim \frac{\rpc P_{\rm CR}}{\epsilon_{\rm CR}} \simeq 400 \bigg(\frac{\gamma_{\rm max}}{10^7} \bigg) \bigg( \frac{\rpc}{1 \ {\rm km}} \bigg) \bigg(\frac{1000 \ {\rm km}}{\rho_c} \bigg).
\end{align}

For each photon (indexed by $i$, with $i = 1, \, 2, \, \dots$) in generation $n$ we compute the mean free path $\ell^{(n)}_i$ to pair production as described above and produce generation $n+1$ pairs and synchrotron photons at the point of absorption. We run five generations of particle production as described in Fig.~\ref{fig:flowchart}. The output of the algorithm gives information about the particle density as a function of position and time within the gap. We use this output to determine locations at which the charge density equals $n_\text{GJ}$, which serve as source points for gap collapse as described in the following subsection. 

{\bf Screening of $\epar$: } The 2+1 dimensional semi-analytic model is constructed by fixing the spatial profile for the un-screened gap $\epar(\vec{r})$, and mimicking the screening behavior induced by pair production processes using damped plane waves -- notice that this is motivated by the qualitative features observed in the simulations of~\cite{Cruz2021}. The physical scales driving the damping and oscillations of the plane waves are inferred using the simplified 1+1 dimensional analytic solutions, as well as the insight gained from the 2.5 dimensional PIC simulations. The procedure followed is outlined below.

We start by defining the physical dimensions of the vacuum gap. The radius of the polar cap $r_{\rm pc}$ is set by the typical footprint produced from a dipolar magnetic field (see Eq.~\eqref{eq:rpc}) and the height of the gap $h$ can be estimated by following the trajectories of $e^\pm$ particles (which travel along magnetic field lines), and determining the point at which pair-production generated a GJ charge density (as described above); this height is approximately given by~\cite{RudermanSutherland1975}
\begin{equation}
h \simeq 50 \, {\rm m} \, \left(\frac{\rho_c}{10^6 \, {\rm cm}} \right)^{2/7} \, \left(\frac{\Omega_{\rm NS}}{1 \, {\rm Hz}} \right)^{-3/7} \, \left(\frac{B_0}{10^{12} \, {\rm G}} \right)^{-4/7} \, ,
\end{equation}
where $\rho_c$ is the curvature of magnetic field lines near the neutron star surface. For a dipolar magnetic field, the radius of curvature is given by~\cite{Timokhin:2018vdn}
\begin{equation}\label{eq:rpc}
    \rho_c \simeq 9.2 \times 10^7 \left(\frac{\theta_{\rm pc}}{\theta} \right) \, \sqrt{P_{\rm NS}} \, {\rm cm} \, ,
\end{equation}
where $\theta$ is the colatitude of the footprint of the magnetic field line, $\theta_{\rm{pc}}$ the colatitude of the polar-cap boundary, and $P_{\rm NS}$ the rotational period of the neutron star. In determining the gap height we set $\theta$ to a constant characteristic value of $\theta_{\rm{pc}} / 2$~\cite{Timokhin:2018vdn}.

The size of $\epar$ may vary within the gap itself, growing linearly along the magnetic axis and decaying off at radial distances near $r_{\rm pc}$. In order to incorporate smooth spatial boundaries we define a gap profile $\psi(r, z)$ such that the maximum unscreened electric field is given by $E_{\parallel, \rm max}(r, z) = E_{\parallel, \rm max} \times \psi(r,z)$, with $E_{\parallel, \rm max} \equiv e \, n_{\rm GJ} \, r_{\rm pc}$. We take the gap profile to be
\begin{equation}
   \psi(r, z) =  \frac{1}{2} \left[ 1 + \tanh\left( \frac{r - r_{\rm pc}}{\delta r} \right) \right] \, \times \left(\frac{h - z}{h} \right) \, ,
\end{equation}
with $\delta r = r_{\rm pc} / 10$. Notice that the radial dependence of $\psi$ is somewhat ad hoc (although similar functional forms have been adopted \eg in~\cite{Cruz2021}). In reality, we simply require a quick and smooth fall-off of the gap near $r_{\rm pc}$, and we expect other types of smooth functional cut-offs to provide comparable results. The linear growth of $\psi$ on the other hand is intended to mimic the inability of the extracted surface current to screen $E_{||}$, the efficiency of which should decrease as one moves away from the surface (see \eg ~\cite{TimokhinArons2013, Philippov2020}). We emphasize that the precise details of this gap profile do not have a large impact on the axion spectra.

The initial conditions of the semi-analytic model are that of a fully grown gap (\ie $\epar = E_{\parallel, \rm max}(r, z)$). At this point, pair production is expected to collapse the gap, sending $\epar \rightarrow 0$; the nonlinear interplay between the motion of the plasma and the response of the electric field, however, induces damped oscillations in $\epar$, with each phase of the oscillation triggering the production of higher generation $e^\pm$ pairs. The damping is expected to be exponential, with a typical timescale roughly given by~\cite{Tolman:2022unu}
\begin{equation}
    t_c \simeq 3.1 \times 10^7 \, \left(\frac{B_0}{10^{12} \, \rm G} \right) \,  \left(\frac{10^5}{\lambda} \right)^{1/2} \,  \left(\frac{0.1 \, {\rm s}}{P_{\rm NS}} \right)^{7/4} \times \frac{m_e}{e E_{\parallel, \rm max}} \, ,
\end{equation}
where $\lambda$ is the charge multiplicity. The frequency of oscillation is set by the local plasma frequency. We assume here that the plasma frequency during the initial response is set by the GJ value (\ie the minimal value required in order to suppress the electric field), and increases by a factor of $100$ as subsequent generations of particles are produced (see \eg~\cite{Tolman:2022unu}). The growth of the oscillation frequency should be slightly less than exponential, since only a subset of each generation's $e^\pm$ pairs  contribute in producing the subsequent generation (at this point, pair production is driven only by the high energy tail of the distribution). As such, we adopt a time-dependent evolution of the oscillation frequency given by
\begin{equation} \label{eq:freqE}
    \tilde{\omega}(t) = \omega_{\rm GJ} \times (1 + \beta e^{\left(t / t_c \right)^\alpha} ) \, .
\end{equation}
Here $\beta \equiv (\hat{\omega}_f - 1) \times e^{-2 \alpha}$, and we set $\alpha = 0.5$ and $\hat{\omega}_f = 100$. In the following section we illustrate that this choice does not have a major impact on the derived constraints.

\begin{figure*}
    \centering
    \includegraphics[scale=0.7]{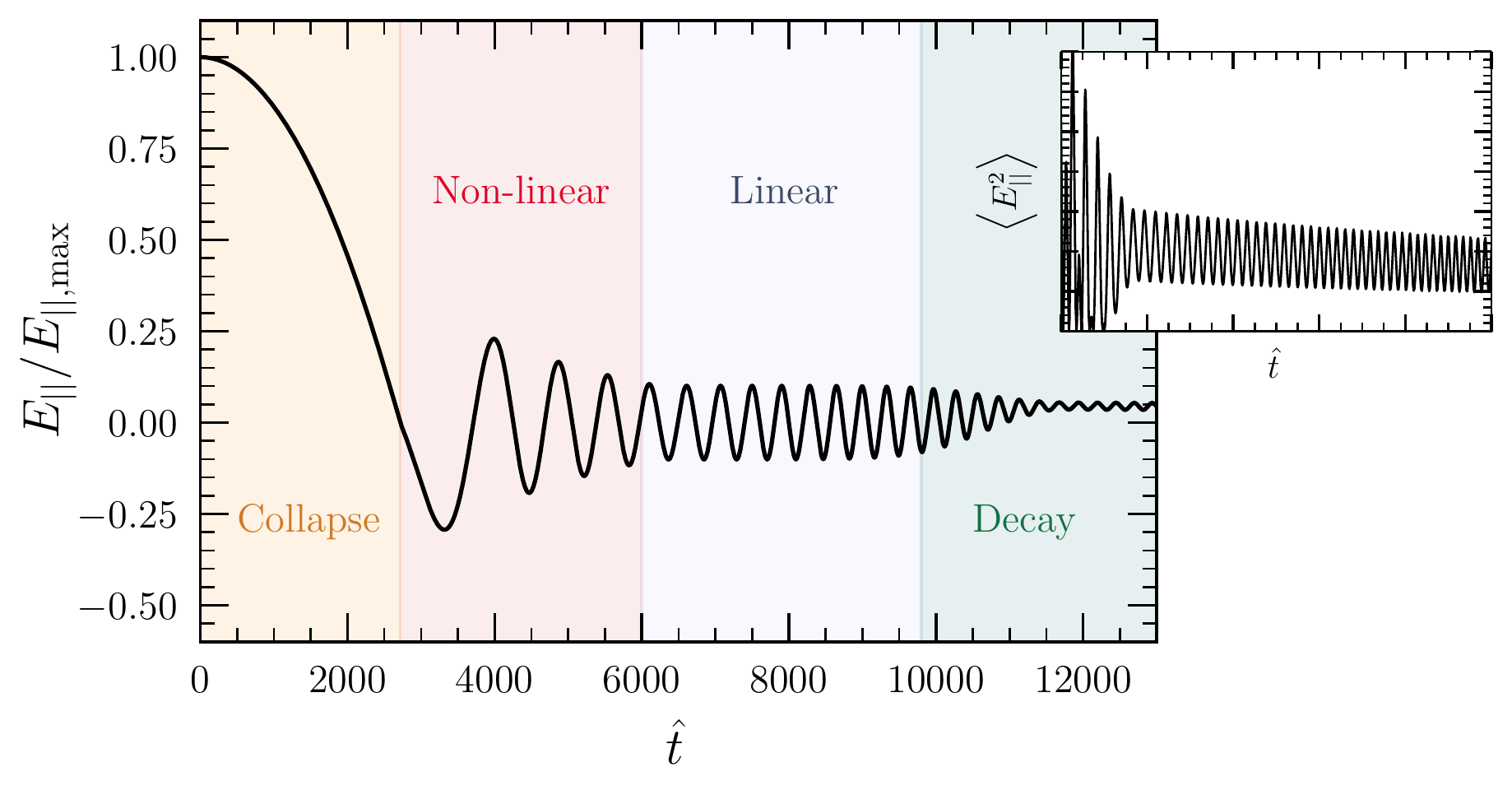}
    \caption{Illustrative evolution of various phases (excluding growth phase) of the evolution of $E_{\parallel}$, following the 1+1 dimensional formalism of~\cite{Tolman:2022unu}. The inset shows a zoom-in of $\left<E_{\parallel}^2 \right>$ during the the decay phase. }
    \label{fig:gap_evol}
\end{figure*}

As an informative illustration, we show in Fig.~\ref{fig:gap_evol}  the temporal evolution of the 1+1 dimensional solution identified in~\cite{Tolman:2022unu} (computed using the same parameters as used in their Fig.~2) as a function of (re-normalized) time $\hat{t}$, and using the frequency evolution in Eq.~\eqref{eq:freqE}. In the figure, we extend the result beyond the nonlinear damping regime, including the linear oscillation and decay regime identified in~\cite{Tolman:2022unu}. We note that the latter two regimes are not significant for this work, as the solution at these late times is only expected to be representative of small localized plasma over-densities located well outside of the polar cap.

The fundamental difficulty in modeling the gap collapse processes arises from the nonlinear response that appears in two dimensions. Particle production itself is inherently inhomogeneous since pair production depends heavily on the radius of curvature of the magnetic field lines, which varies across the gap. The response of the electromagnetic fields to the inhomogeneity in the particle production further induces nonlinear dependencies, which significantly complicate the evolution of $\epar$. Here,  we attempt to capture the initial inhomogeneity in the pair formation front by accelerating test particles and tracing the radiation and initial pair production events (see section on pair cascade above); this procedure neglects the back-reaction on the electromagnetic fields, and thus is only valid when $\epar$ is large. 

We model the subsequent nonlinear screening of $\epar$ by drawing well-isolated samples from the pair formation front, and inducing two-dimensional damped plane waves that propagate outward -- we refer to these samples as `pair production seeds', with coordinates denoted by $(r_{{\rm seed}, i}, \, z_{{\rm seed}, i})$. Note that similar behavior can be seen in the simulations of~\cite{Fabio2021}. We approximate the evolution of $\epar$ using the Ansatz
\begin{equation}\label{eq:epar_samod}
    \epar(r, z, t) = E_{\parallel, \rm max} \times \psi(r, z) \times \prod_{i = 1}^{N_{\rm pts}} \, \phi_{{\rm pp}, i}(r, z, t) \, ,
\end{equation}
where the product runs over each of the sampled seeds. Our intent with this model is to capture the typical temporal and spatial variations of $\epar$.  We will validate it below by comparison with PIC simulation results. The response of each seed is given by
\begin{equation}\label{eq:seeds}
    \phi_{{\rm pp}, i}(r, z, t) = \begin{cases} 
    \cos(\tilde{\omega} \, \tilde{t}_i - k \cdot d_{{\rm pp}, i}) \, e^{- \tilde{t}_i / t_c}  \, \hspace{.5cm} & {\rm{if} \, } \tilde{t}_i \geq 0 \, , \\
    1  \, \hspace{.2cm} & {\rm{if} \, } \tilde{t}_i < 0 \, .
    \end{cases}
\end{equation}
We take $k = \tilde{\omega}$ (\ie assume it travels at the speed of light), and the functional dependencies on spacetime are understood to be implicit. We have introduced in Eq.~\eqref{eq:seeds} the notion of an `effective time' $\tilde{t}_i$ which captures the relative spacetime dependent response of $\epar$ (in other words, this captures the relative delay that would arise in the event that gap collapse occurs locally and expands outwards, as seen in~\cite{Fabio2021}); this is given by
\begin{equation}
    \tilde{t}_i(r,z,t) \equiv t - t_{{\rm pp}, i} - d_{{\rm pp}, i}(r, z, t) \, ,
\end{equation}
where $t$ is the absolute time, $t_{{\rm pp}, i}$ is the initial time at which particle production event $i$ was initiated, and $d_{{\rm pp}, i}$ is the relative distance from the center of particle production event $i$ to the position $(r,z)$. Since the electric field will push the newly produced plasma out of the gap, we adopt a drift in the initial seed location such that
\begin{equation}
    z_{{\rm pp}, i} = \begin{cases}
    z_{{\rm seed}, i}  \, \hspace{.5cm} & {\rm{if} \, } (t - t_{{\rm pp}, i}) < t_{\rm delay} \, , \\
    z_{{\rm seed}, i} + (t - t_{{\rm pp}, i} - t_{\rm delay})  \, \hspace{.5cm} & {\rm otherwise} \, .
    \end{cases}
\end{equation}
The factor $t_{\rm delay}$ is set by the minimum time required for the plane waves to collapse the gap at all radii, which is also roughly the time over which the pair formation front forms, \ie ${\rm max}(t_{{\rm pp}, i}) - {\rm min}(t_{{\rm pp}, i})$ (note that the inclusion of $t_{\rm delay}$ is necessary to ensure that the gap collapses). We run each semi-analytic model for a total time set by $t_{\rm max} = 2 \times {\rm max}(t_{{\rm pp}, i})$, which is sufficiently long to observe the gap collapse and the subsequent outward drift of the seed points; moreover, this number agrees quite well with the expected gap collapse period of typical neutron stars, which is around $10^{-9} - 10^{-6} \, \rm s$~\cite{Cruz:2022zyp}.

\begin{figure*}
    \centering
    \includegraphics[width=\textwidth]{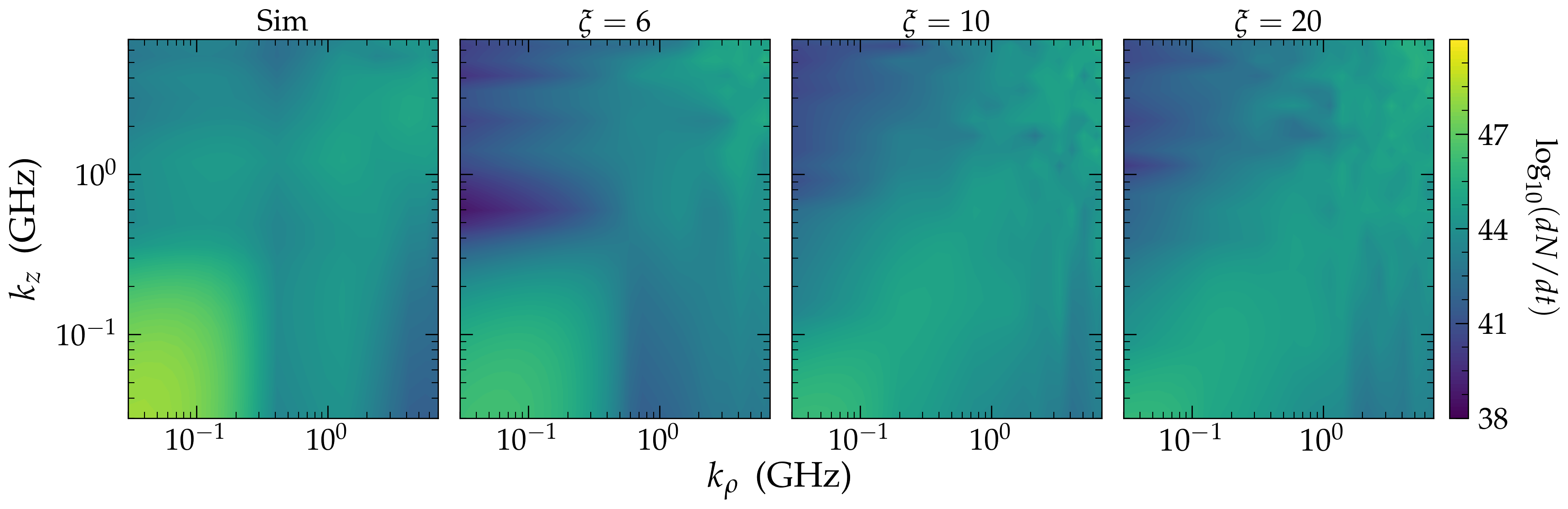}
    \caption{Number of axions produced per unit time predicted by the simulation (left panel) and semi-analytic model (right three panels) for various extrapolation factors $\xi$, ranging from 6 to 20. The spectra are produced for pulsar B1737-30 (see Table~\ref{tab:pulsar}), an axion mass of $10^{-8} \eV$, and an axion-photon coupling of $g_{a\gamma\gamma} = 10^{-11} \, {\rm GeV}^{-1}$. }
    \label{fig:rescalingF}
\end{figure*}

In order to be able to perform the full FT across the momentum range of interest one must have sufficient spatial and temporal resolution. Specifically, we require a spatial resolution at the level of the GJ oscillation wavelength (which is well-beyond the resolution needed to resolve momentum modes $k \sim \mathcal{O}(10 \GHz)$), and temporal resolution given by
\begin{equation}
    \delta t = {\rm min}\left(\frac{2\pi}{\omega_{\rm GJ}} , \frac{2\pi}{\omega_{\rm max}} \right) \, ,
\end{equation}
where $\omega_{\rm max}$ is the maximum axion energy computed. For most of the pulsars and masses of interest, storing the gridded evolution of the gap at this resolution requires $\mathcal{O}(100 \, \rm Gbs)$ to $\mathcal{O}({\rm Tbs})$ of memory. 

In order to make these computations more feasible, we perform the semi-analytic calculations with a slightly enhanced value of the GJ scale $\tilde{\lambda}_{\rm GJ} = \xi \times \lambda_{\rm GJ}$, with $\xi$ taken to be 6 (roughly the smallest value which we can achieve given current resources). Similarly to the case of the PIC simulation, this modification effectively inflates small scales and thus mis-places the power of the FT in smaller $k$-modes (here, we resolve around 4.5 orders of magnitude in scale, but require around 5 -- thus the effect is rather minimal). As before, we attempt to correct for this by first computing $\tilde{\jmath}$ at a series of log-spaced momentum modes $(\tilde{k}_{\rm min}, \cdots, \tilde{k}_{\rm max})$, and subsequently extrapolating the momentum vector $k_i = (\xi)^{i/n} \, \tilde{k}_i$ before calculating the differential rate in Eq.~\eqref{eqn:diffrate}. To illustrate the sensitivity to this extrapolation factor, we plot in Fig.~\ref{fig:rescalingF} the axion spectrum predicted using $\xi = 6, \, 10,$ and $20$ for pulsar B1737-30 and an axion mass of $10^{-8}~$eV (the axion-photon coupling is set to $g_{a\gamma\gamma} = 10^{-11} \, {\rm GeV^{-1}})$. In the left panel of Fig.~\ref{fig:rescalingF}  we also show for comparison the corresponding spectrum produced by the simulation. One can see that the relative difference between the simulation and semi-analytic model greatly exceeds the small differences that appear as one changes the re-scaling factor. All runs are computed at 35 log-spaced $k$-values in both $k_\rho$ and $k_z$. In SM Section~\ref{sec:uncert}, we return to the effect of this re-scaling factor and show that taking a value of $\xi = 10$ makes only minimal difference in the derived limits, thus justifying the adopted re-scaling factor used for the fiducial analysis.

\section{Computing the radio flux}\label{app:radio}
From the spectra computed using the PIC simulation and the semi-analytic model, we can find the initial spectrum of axions produced in the gap. With the initial axion spectrum in hand, one can subsequently calculate the radio flux generated via resonant axion-photon transitions through: $(i)$ carefully following the trajectories of all axions created in the polar caps, $(ii)$ identifying the resonant level crossings, $(iii)$ computing the conversion probabilities from axions into photons, and $(iv)$ propagating the produced photons through the dispersive magnetosphere to find their direction and energy at asymptotic infinity. The ray-tracing code premiered in~\cite{Witte2021} was developed to treat the resonant conversion and propagation of radio flux arising from axion dark matter around neutron stars; we have modified this code to include axion propagation in a Schwarzschild background and incorporated a generalization of the mixing which is valid for (ultra-)relativistic axions. Both of these new features, as well as other relevant steps in computing the axion-induced radio flux, are outlined below. 

\subsection{Axion Production Rate}
In this section we review the derivation of Eq.~\eqref{eqn:diffrate} from Eq.~\eqref{eqn:KG}. We follow the discussion in~\cite{peskin2018introduction}. In the particular case of electromagnetic production of axions the classical source is given by $j(x) = -\gagg (\vec{E} \cdot \vec{B})(x)$. The average production rate can be calculated assuming the source is non-vanishing only over a time period $T$, which in our scenario corresponds to the gap collapse time. In the asymptotic past the axion field can then be expanded as 
\begin{align}
        a_0(x) = \displaystyle\int \frac{d^3 k}{(2\pi)^3 \sqrt{2 \omega(\vec{k})}} \left( a_{\vec{k}} e^{-i k \cdot x} + a_{\vec{k}}^\dagger e^{i k \cdot x} \right) \, ,
\end{align}
where $a_{\vec{k}}^\dagger \left( a_{\vec{k}} \right)$ is the creation (annihilation) operator for the axion and $\omega(\vec{k}) = \sqrt{k^2 + m_a^2}$. This can be thought of as the homogeneous solution to Eq.~\eqref{eqn:KG} or, equivalently, the so-called `in' state. The effect of the source, after it has been turned off, is to shift the creation and annihilation operators
\begin{align} \label{eqn:BogoTrans}
    a_{\vec{k}} \to a_{\vec{k}} + \frac{i}{\sqrt{2\omega(\vec{k})}} \tilde{\jmath}(\vec{k}), \quad a^\dagger_{\vec{k}} \to a^\dagger_{\vec{k}} - \frac{i}{\sqrt{2\omega(\vec{k})}}\tilde{\jmath}(-\vec{k}), \quad     \tilde{\jmath}(\vec{k}) = \displaystyle\int d^4 x \, e^{i k \cdot x} j(x) \, ,
\end{align}
where $\tilde{\jmath}(\vec{k})$ is the Fourier transform of $j(x)$. It follows from Eq.~\eqref{eqn:BogoTrans} that the average rate of particle production is given by
\begin{align}
    \frac{d \dot{N}}{d^3 k} = \frac{\Braket{0|a_{\vec{k}}^\dagger a_{\vec{k}}|0}}{(2\pi)^3 \, T} = \frac{\left| \tilde{\jmath}(\vec{k}) \right|^2}{2 (2\pi)^3 \, \omega(\vec{k}) T} \, ,
\end{align}
where $\ket{0}$ is the asymptotic vacuum state. 

\subsection{Axion propagation}
Since axions are produced close to the neutron star (where gravity is strong), we propagate axions assuming they follow Schwarzschild geodesics\footnote{This assumption is of course not true inside the neutron star itself. While our signal is not significantly affected by through-passing axions, we choose to re-scale the neutron star mass by the fraction of mass contained in a radius $r$ (assuming a constant density profile) in order to avoid singularities. In the future one could improve upon this by adopting the interior  Schwarzschild metric and taking the density profile under a fixed equation of state. }. The relevant equations describing the axion trajectories are
\begin{gather}
    \frac{d^2r}{d\tau^2} = -\frac{GM_{\rm NS}}{r^2} + \frac{L^2}{m_a^2 r^3} - \frac{3 G M_{\rm NS} L^2}{m_a^2 r^4} \, , \\
    \frac{d^2\phi}{d\tau^2} = - \frac{2}{r} \frac{dr}{d\tau} \frac{d\phi}{d\tau} \, .
\end{gather}
Here we have chosen to work in polar coordinates, with $\tau$ specifying the proper time and $G$ being the gravitational constant. It is important to note that both the angular momentum and total energy are constant along the trajectories and given by
\begin{gather}
    L = m_a v_{\perp} \gamma r \, , \\
    E = m_a \gamma \sqrt{1 - \frac{r_s}{r}} \, ,
\end{gather}
where $\gamma$ is the standard Lorentz factor, $v_{\perp}$ is the perpendicular component of the local particle velocity and $r_s$ is the Schwarzschild radius of the neutron star.

The axions are propagated in just two dimensions, as their trajectories are inherently symmetric about the magnetic axis of the neutron star. The locations of the conversion points, however, are not. Consequently, we project each calculated axion trajectory in 70 different randomly sampled azimuthal directions when computing the conversion points and probabilities. This amount of samples was chosen to maximize stability and accuracy in the final results while maintaining computational tractability. After ray-tracing, we average over the rotational period to compute the expected radio flux.

\subsection{Axion-photon conversion}
In its most general form, the axion-photon transition probability is given by the Landau-Zener equation
\begin{equation}\label{eq:lz}
P_{a\rightarrow \gamma} = 1 - e^{-\Gamma} \, .
\end{equation}
This properly describes conversion in the adiabatic (\ie large coupling and/or large magnetic field) regime, assuming that the length scale over which the mixing takes place is sufficiently small with respect to variations in the magnetic field and plasma density\footnote{Recent work in~\cite{Carenza:2023nck} has shown that Eq.~\ref{eq:lz} breaks down when the conversion is adiabatic and when these scales are comparable. For small axion masses, we do find that some photons are sourced from resonances in which Eq.~\ref{eq:lz} does not hold; however, removing these photons entirely leads to a modest correction to the constraints on the order of $\lesssim 10\%$ for axion mass $m_a = 10^{-9} \, {\rm eV}$ and $\ll 1\%$ for $m_a \geq 10^{-8} \, {\rm eV}$, implying our constraints are robust to this uncertainty. We caution that lower mass axions are likely to be more severely affected by this problem, and thus a careful treatment would be required in this regime.}. Details in calculating the form of the adiabaticity parameter $\Gamma$ from the electrodynamic equations of motion can be found in~\cite{Witte2021}. Here, we generalize the aforementioned analysis to include: $(i)$ relativistic mixing and $(ii)$ the latest approximation of the axion-photon conversion probability from~\cite{millar2021axionphotonUPDATED}. 

Resonant axion-photon transitions occur when $k_a \simeq k_\gamma$~\cite{Witte2021}, where the axion momentum is given by
\begin{equation}
    k_\gamma = \sqrt{\frac{\omega^2 - \omega_p^2}{1 - \frac{\omega_p^2}{\omega^2}\cos^2\theta}} \, .
\end{equation}
As in the main text, $\omega$ is the axion/photon energy, $\omega_p$ is the plasma frequency, and $\theta$ is the angle between the axion momentum and the external magnetic field $\vec{B}$. In the non-relativistic limit, this resonance happens at $m_a \sim \omega_p$; the generalized condition (which is also applicable in the relativistic regime), however, has the form
\begin{equation}
    \omega_p^2 \simeq \frac{m_a^2 \omega^2}{m_a^2 \cos^2\theta + \omega^2 \sin^2\theta} \, .
\end{equation}

Recently, Ref.~\cite{millar2021axionphotonUPDATED} pointed out that off-diagonal derivatives (\eg $\partial_{\hat{k}_{\parallel}, \hat{k}_{\perp}}  E_\perp$) in the axion-photon mixing equations modify the conversion probability. In particular, this contribution shows that the group velocity of the sourced electromagnetic mode travels in the direction $\hat{s}$, which is defined through its differential
\begin{equation}\label{eq:s_diff}
\partial_s \equiv \partial_{\hat{k}_{\parallel}} - (\omega_p^2 \Delta \cos\theta / \omega^2) \partial_{\hat{k}_{\perp}} \, .
\end{equation}
Here $\hat{k}_{\parallel}$ and $\hat{k}_\perp$ are the parallel and perpendicular directions to the axion momentum respectively (note that we have defined $\hat{k}_\perp$ to lie in direction of $\hat{B}$, which generates a sign difference with respect to~\cite{millar2021axionphotonUPDATED}), and
\begin{equation}
    \Delta = \frac{\sin\theta}{1 - \frac{\omega_p^2}{\omega^2} \cos^2\theta} \, .
\end{equation}
Since the axion-photon conversion probability depends on change of the photon momentum along the direction of the group velocity, one must derive the correction to the adiabaticity parameter arising from these off-diagonal derivatives; the result is given by~\cite{millar2021axionphotonUPDATED}
\begin{equation}\label{eq:conv}
    \Gamma = \frac{\pi}{2} \Big(1 + \frac{\omega_p^4 \Delta^2 \cos^2\theta}{\omega^4}\Big) \Big(\frac{\omega g_{a\gamma\gamma} B \Delta}{k_a}\Big)^2 \frac{1}{\lvert \partial_s k_\gamma \rvert} \, .
\end{equation}

\subsection{A brief comment on adiabatic axion-photon conversion}
It was recently pointed out in~\cite{Foster:2022fxn} that axion-photon conversion in the magnetospheres of highly magnetized neutron stars can become extremely efficient, generating order one conversion probabilities. In the context of axion dark matter (in which axions are assumed to fall onto the neutron star), this can lead to an exponential suppression of the radio flux -- this is a consequence of the fact that each axion is expected to pass an even number of resonances, and thus the generation of an observable radio signal requires a produced photon to survive a level-crossing during escape from the magnetosphere (and $P_{\gamma \rightarrow \gamma} = 1 - P_{a\rightarrow \gamma} \ll 1$ when the conversion probability is large). We emphasize here that the localized axion production in the polar caps naturally evades this suppression as these axions typically undergo an odd number of level crossings, making this a highly complementary probe of axions in the large coupling limit.

\begin{figure*}
    \centering
    \includegraphics[width=.49\textwidth]{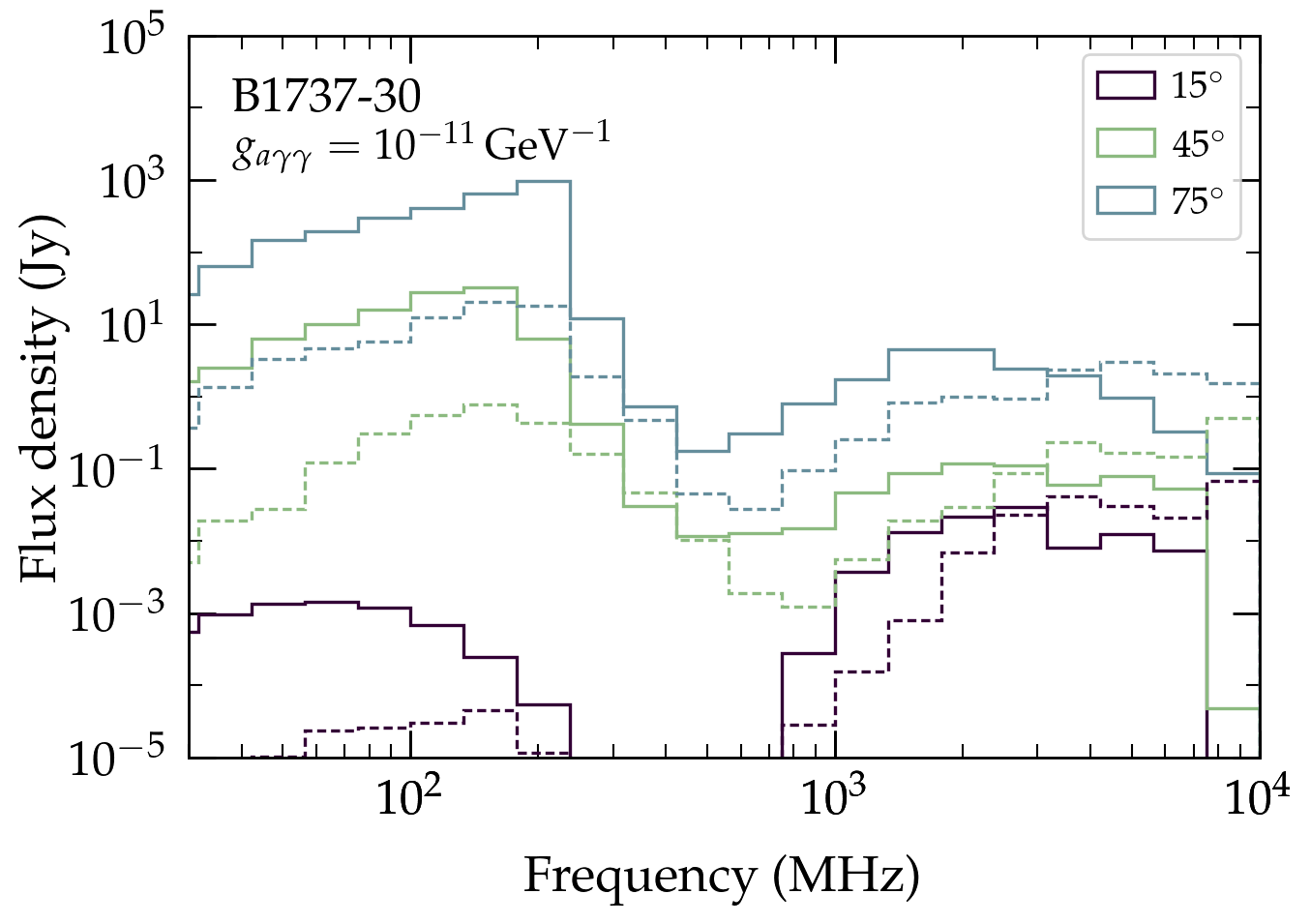}
    \includegraphics[width=.49\textwidth]{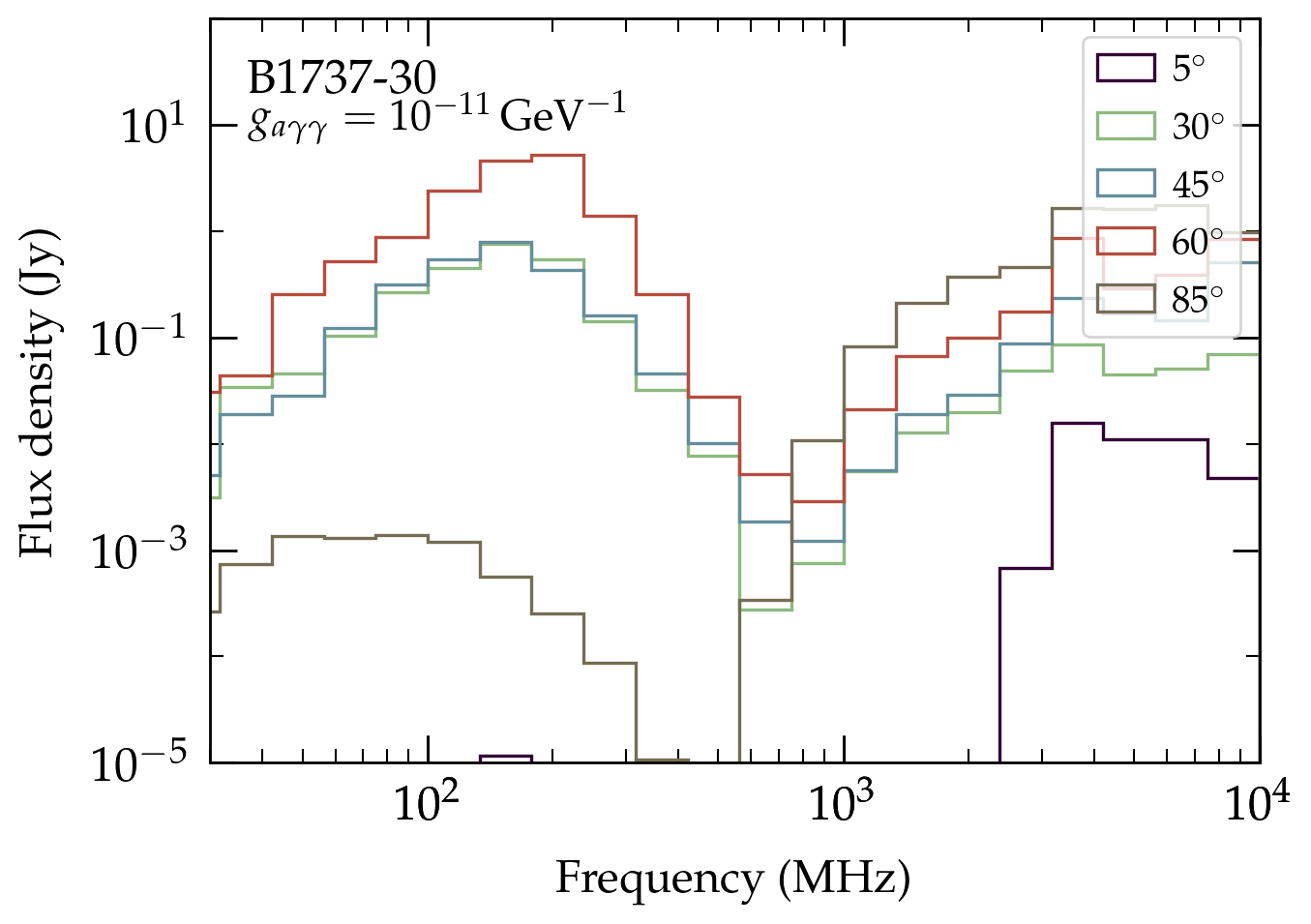}
    \caption{Left: Comparison of photon spectra produced by pulsar B1737-30 using the PIC simulation (solid) and the semi-analytic model (dashed). Results are shown for three different viewing angles. Right: Photon spectra produced by pulsar B1737-30 using the semi-analytic model over a wider range of viewing angles. All results assume an axion mass of $10^{-8} \, {\rm eV}$ and an axion photon coupling $g_{a\gamma\gamma} =  10^{-11} \, {\rm GeV}^{-1}$.}
    \label{fig:spectrum}
\end{figure*}

\subsection{Example photon spectra}
Before providing an illustration of the photon spectra produced through our models, we reiterate the full procedure used to obtain these spectra. This procedure includes:
\begin{itemize}
    \item We first compute the axion production rate across a range of momentum bins using either the re-scaled PIC simulation or the semi-analytic model.
    \item We then use these sampled momenta as initial conditions for our ray-tracing algorithm. Axions are traced away from the gap, and their resonances along the trajectory are identified.
    \item For each resonance, we compute the probability of sourcing a radio photon (which depends on the conversion probability at the resonance, as well as the survival probability of the axion at previous resonances) and ray-trace the photon to large distances.
    \item After all photons have been traced, we determine the relative viewing angle of Earth, $\theta_\oplus$, and compute the average radio flux by summing over the weighted contributions of photons with final positions $\theta \sim \theta_\oplus \pm \epsilon$ (with $\epsilon$ taken to be very small and with photon frequencies falling in the desired bin $[\nu_1, \nu_2]$. 
\end{itemize}

We provide an illustrate of the photon spectra produced by pulsar B1737-30 in Fig.~\ref{fig:spectrum}, where the left panel shows a comparison between the predictions of the simulation and the semi-analytic model (at three viewing angles), while the right panel displays the observed flux density at five viewing angles (computed using only the semi-analytic model). In both cases, the axion mass is set to $m_a = 10^{-8} \, {\rm eV}$ and the axion-photon coupling to $g_{a\gamma\gamma} = 10^{-11} \, {\rm GeV}^{-1}$. Fig.~\ref{fig:spectrum} highlights two important points: there is reasonable agreement between the spectrum predicted by the PIC simulation and that of the semi-analytic model, and the observed flux density depends rather strongly on the observed viewing angle. 

\begin{table*}[!ht]
\setlength\extrarowheight{5pt}
\footnotesize
\begin{tabular}{|c|c|c|c|c|c|c|c|c|c|c|}
\hline
Name & Period (s) & $B_0$ (G) & $\theta_m$ (deg) & $\beta$ (deg) & $S_{25-80}$ (mJy) & $S_{50-80}$ (mJy) & $S_{408}$ (mJy) & $S_{1400}$ (mJy) & $S_{8600}$ (mJy) & Dist. (kpc) \\ \hline \hline 
B1822-09 & 0.77 & 6.42E12 & 95 & -7  & - & 2502 $\pm 1251$ & $35.5 \pm 2.5$ & $10.2 \pm 20 $ & - & 0.3$^{+0.7}_{-0.2}$ \\ \hline 
B2045-16 & 1.96 & 4.69E12 & 36 & 1.1 &  - & - & 115 $\pm 26$  & $22 \pm 44$ & $0.3 \pm 0.04$ & 0.95$^{+0.02}_{-0.03}$ \\ \hline
B0450-18 & 0.55 & 1.80E12 & 24 & 4  & -  & - & 82.4 $\pm 10$ & $16.8 \pm 34$ & - & 0.4$^{+0.2}_{-0.1}$ \\ \hline 
B0329+54 & 0.71 & 1.22E12 & 30 & 2.1  & - & 1841$\pm 921$ & $1512 \pm 175$ & $203 \pm 57$ & - & 1.0$^{+0.1}_{-0.1}$ \\ \hline
B0450+55 & 0.34 & 9.10E11 & 32 & 3.3  & 124$\pm 62$  & - & 59.0 $\pm 9$ & $13.0 \pm 3$ & - & 1.18$^{+0.07}_{-0.05}$ \\ \hline 
B0656+14 & 0.38 & 4.65e12 & 30 & 8.2  & - & - & 6.5$\pm 0.6$ & $2.7 \pm 2$ & $0.96 \pm 0.11$ & 0.29$^{+0.03}_{-0.03}$ \\ \hline 
B0919+06 & 0.43 & 2.46e12 & 48 & 4.8  & -  & 550 $\pm 275$ & $52.2 \pm 6.4$ & $10 \pm 3$ & - & 1.1$^{+0.2}_{-0.1}$ \\ \hline 
B1508+55 & 0.74 & 1.95e12 & 45 & -2.7 & - & 943 $\pm 471$ & 114$\pm 5$ & $8 \pm 1$ & - & 2.1$^{+0.1}_{-0.1}$ \\ \hline 
B1738-08 & 2.04 & 2.18E12 & 26 & 1.7 & - & - & 29.4$\pm 7.7$ & $1.4 \pm 4.0$ & - &  0.2$^{+0.2 \, \dagger}_{-0.2}$ \\ \hline 
B1818-04 & 0.60 & 1.97E12 & 65 & 3.5 & - & - & 156$\pm 6$ & $10.07 \pm 2.0$ & -  & 0.6$^{+0.03}_{-0.03}$ \\ \hline 
B1917+00 & 1.27 & 3.16E12 & 81 & 1.3 & - & -  & 15.6$\pm 1.3$ & $0.8 \pm 2.0$ & - & 3.6$^{+0.5 \, \dagger}_{-0.5}$ \\ \hline 
B1919+14 & 0.62 & 1.88E12 & 26 & -6.4 & -  & - & $<$5.2 & $0.68 \pm 8$ & - & $3.4^{+0.55 \, \dagger}_{-0.55}$ \\ \hline 
B1919+21 & 1.34 & 1.36E12 & 45 & -3.7  & $1586 \pm 793$ & - & 56.5 $\pm 8.2$ & $18.8 \pm 38$ & - & 0.3$^{+0.8}_{-0.2}$\\ \hline 
B1920+21 & 1.08 & 3.00E12 & 44 & 1.1  & -  & -  & 29.9$\pm 1.2$ & $1.4 \pm 2.0$ & - & 4.0$^{+2.0}_{-2.0}$ \\ \hline 
B2224+65 & 0.68 & 2.60E12 & 16 & 3.4  & - & $293 \pm 146$ & $21.9 \pm 2.4$ & $2 \pm 4$ & - & 0.9$^{+0.1}_{-0.1}$ \\ \hline 
B2319+60 & 2.26 & 4.03E12 & 18 & 2.2 & - & - & $36.1 \pm 4.9$ & $12 \pm 1$ & - & 2.7$^{+1.2}_{-0.9}$ \\ \hline 
B1910+20 & 2.23 & 4.82E12 & 29 & 1.5 & - & -  & $5.7 \pm 1.1$ & $0.22 \pm 3$ & - & 3.6$^{+0.3 \, \dagger}_{-0.3}$\\ \hline 
B0402+61 & 0.59 & 1.84E12 & 83 & 2.2 & - & - &  14.6$\pm 1.3$& $2.8 \pm 2$ & - & 1.95$^{+0.17 \, \dagger}_{-0.17}$ \\ \hline 
B1039-19 & 1.39 & 1.16E12 & 31 & 1.7  & - & - & 27.7$\pm 5.5$ & $0.62 \pm 7$ & - & 2.0$^{+0.5 \, \dagger}_{-0.5}$ \\ \hline 
B1237+25 & 1.38 & 1.17E12 & 53 & 0  & $102 \pm 51$ & - & $110 \pm 33$ & $23.2 \pm 47.0$ & $0.31 \pm 0.07$ & 0.84$^{+0.06}_{-0.05}$ \\ \hline 
B1737+13 & 0.80 & 1.09E12 & 41 & 1.9  & - & $131 \pm 66$ & $23.9 \pm 1.8$ & $8.9 \pm 5$ & -  &  4.2$^{+4.2 \, \dagger}_{-4.2}$ \\ \hline 
B1737-30 & 0.61 & 1.70E13 & 58 & 10.9  & - & - & $24.6 \pm 0.1$ & $8.9 \pm 5$ & $1.21 \pm 0.07$  &  0.4$^{+1.7}_{-0.3}$ \\ \hline
B0833-45 & 0.089 & 3.38E12 & 90 & 12.0  & - & - & $5000 \pm 0.1$ & $1050 \pm 60$ & -  &  0.28$^{+0.02}_{-0.02}$ \\ \hline
B0923-58 & 0.74 & 1.93E12 & 19 & 3.0  & - & - & $22 \pm 0.1$ & $21 \pm 6$ & -  &  0.107$^{+0.1 \, \dagger}_{-0.1}$ \\ \hline
B1742-30 & 0.367 & 2.0E12 & 24 & 6.4  & - & - & $66 \pm 6$ & $21 \pm 1$ & $1.55 \pm 0.05$  &  0.2$^{+1.1}_{-0.2}$ \\ \hline
B1749-28 & 0.56 & 2.16E12 & 42 & 2.9  & - & - & $1100 \pm 100$ & $47.8 \pm 96$ & $0.3 \pm 0.05$  &  0.2$^{+1.1}_{-0.1}$ \\ \hline
B2334+61 & 0.495 & 9.91E12 & 33 & 3.5  & - & - & $10.0 \pm 2$ & $1.4 \pm 3$ & - &  0.7$^{+0.1}_{-0.1}$ \\ \hline
\end{tabular}
\caption{\label{tab:pulsar}List of pulsars used in this work. For each pulsar the columns denote: name, rotational period (in seconds), inferred dipolar surface magnetic field $B_0$ (in Gauss), the angle $\theta_m$ between the magnetic- and rotational axis (in degrees), the angle $\beta$ between the magnetic axis and the observer at closest approach (in degrees), the flux density between $25-80$ MHz~\cite{bondonneau2020census}, $50-80 \, \rm MHz$~\cite{bondonneau2020census}, $408 \pm 4 \, \rm MHz$~\cite{lorimer1995multifrequency}, $1.4 \pm 0.032 \, \rm GHz$~\cite{lorimer1995multifrequency}, and $8.6 \pm 0.8 \, \rm GHz$~\cite{Zhao:2017tzv} (all in mJy), and the distance (in kpc). The inferred geometric angles are obtained from~\cite{rankin1993toward,gupta2003understanding,rankin2022radio}. Distance measurements and errors are taken from~\cite{Verbiest:2012kh,yao2017new,Brownsberger:2014yya,price2021comparison}; following the ATNF procedure, we prioritize measurements not inferred using the dispersion measure and Galactic electron models. In some cases (marked with a $\dagger$), the Galactic coordinates and dispersion measure are used to infer the distance utilizing the online tools from~\cite{price2021comparison}. Should the various Galactic models agree, we take the mean and standard deviation of these inferred distances; if they instead differ by more than a factor of 2, we defer to the more recent model of~\cite{yao2017new} and place a 100$\%$ error on the distance.
}
\end{table*}

\section{Pulsar Data}\label{app:pulsar}
In order to constrain the existence of axions, we compare the predicted radio flux with observations from 27 nearby pulsars. We have limited ourselves to this small carefully selected sub-sample in order to make the analysis computationally feasible. Our selection is largely based on two fundamental criteria: we want pulsars with large magnetic fields and a well-measured orientation (meaning that both the misalignment angle and the pulsar's orientation relative to Earth are known). The former is particularly important as  the production of axions scales with $\propto B_0^4$ and the resonant conversion process scales with $\propto B(\vec{r}_c)^2$ ($\vec{r}_c$ being the point of resonant conversion). We have chosen to apply the geometrical constraint as the axion-induced radio flux varies rather strongly with viewing angle (see \eg Fig.~\ref{fig:spectrum}), and knowledge of the geometry evades the need for marginalization. It is worth mentioning that a broader study over the entire pulsar population would naturally marginalize over the geometry, and could yield stronger constraints than those obtained here; currently, such a study is computationally prohibitive, however we hope that future improvements will make this manageable. 

The radio observations were obtained using the LOFAR telescope~\cite{bondonneau2020census}, the Lovell telescope at Jordell Bank~\cite{lorimer1995multifrequency}, and the Shanghai Tian Ma Radio Telescope~\cite{Zhao:2017tzv}, and span a wide range of frequencies from $25 \, \rm MHz$ to $8.6 \, \rm GHz$\footnote{The low-frequency spectrum of radio pulsars can, in some cases, be suppressed by free-free absorption~\cite{Jankowski:2017yje}. For each of the pulsars of interest, we use the observed dispersion measure to infer the average line of sight electron number density, and subsequently use this quantity to estimate the optical depth. For all pulsars of interest we find that the optical depth $\tau \ll 1$, suggesting free-free absorption can be neglected.}. The properties of each of the 27 pulsars and the observed flux densities are listed for reference in Table~\ref{tab:pulsar}. Below we outline the analysis used to derive upper limits on the axion-photon coupling, and highlight the impact of various systematic uncertainties in the modeling of the axion signal.

\section{Profile Likelihood Analysis}
We utilize the profile likelihood ratio to derive a one-sided upper limit on $g_{a\gamma\gamma}$. In general, this is accomplished by defining a test statistic~\cite{WilksLikelihood, WaldLikelihood, Cowan:2010js}
\begin{equation}
    {\rm TS}(\mu) = \begin{cases} - 2 \, \textrm{ln} \frac{\mathcal{L}(\mu, \hat{\hat{\vec{\theta}}})}{\mathcal{L}(\hat{\mu}, \hat{\vec{\theta}} \, )} \, & \hspace{.3cm} {\rm if} \, \hat{\mu} \leq \mu \, , \\ 
   0 & \hspace{.3cm} {\rm if} \, \hat{\mu} > \mu \, .
    \end{cases}
\end{equation}
Here $\mu$ is the parameter of interest (in our case, $g_{a\gamma\gamma}$), $\vec{\theta}$ denotes the nuisance parameters (which here include the pulsar distances and the intrinsic radio flux from each pulsar at each frequency), ($\hat{\mu}, \hat{\theta})$ are the maximum likelihood estimators, and $\hat{\hat{\theta}}$ is the conditional maximum likelihood estimator for a given value of $\mu$. 

We adopt Gaussian likelihoods for the flux density and pulsar distances. The full likelihood is given by
\begin{equation}\label{eq:likelihood}
    \mathcal{L}(g_{a\gamma\gamma}, \vec{\theta}) = \prod_i^{N_{\rm pul}} \prod_j^{N_{\rm bin}} e^{-(S_{\textrm{obs}, ij} - S_{\textrm{pred}, ij})^2 / 2 \sigma_{S, ij}^2} \times e^{-(d_{i} - \mu_{d, i})^2 / 2 \sigma_{d, i}^2} \, ,
\end{equation}
where the product is taken over each pulsar and each frequency bin for which observations are available (see Table~\ref{tab:pulsar}). Some important comments regarding this equation:
\begin{itemize}
    \item The predicted flux density has two components, one due to the predicted axion signal and another that is intrinsic to the respective pulsar, \ie $S_{{\rm pred}, ij} = S_{{\rm axion}, ij} + S_{{\rm pulsar}, ij}$. Owing to the fact that we do not currently have a firm understanding of the intrinsic radio flux produced by pulsars, we treat the parameters $ S_{{\rm pulsar}, ij}$ as nuisance parameters. This implies that our test statistic is zero, unless the predicted axion flux in at least one observing bin of one pulsar exceeds the observed flux. This procedure ensures that our constraints remain conservative.
    \item All observed flux densities and mean distances, as well as the variances in both of these quantities, are reported in Table~\ref{tab:pulsar}. Note that the variance in the distance is generally asymmetric, meaning that the second exponential in Eq.~\eqref{eq:likelihood} is actually an asymmetric Gaussian.
    \item The pulsar distances show up explicitly in the likelihood, but also enter through the predicted flux densities.
\end{itemize}
With the likelihood in hand, we compute our test statistic at set couplings and derive an upper limit on $g_{a\gamma\gamma}$ by identifying the value at which $\textrm{TS} = 2.71$. This corresponds to the $95\%$ upper limit~\cite{Cowan:2010js}.

\begin{figure*}
    \centering
    \includegraphics[width=.48\textwidth]{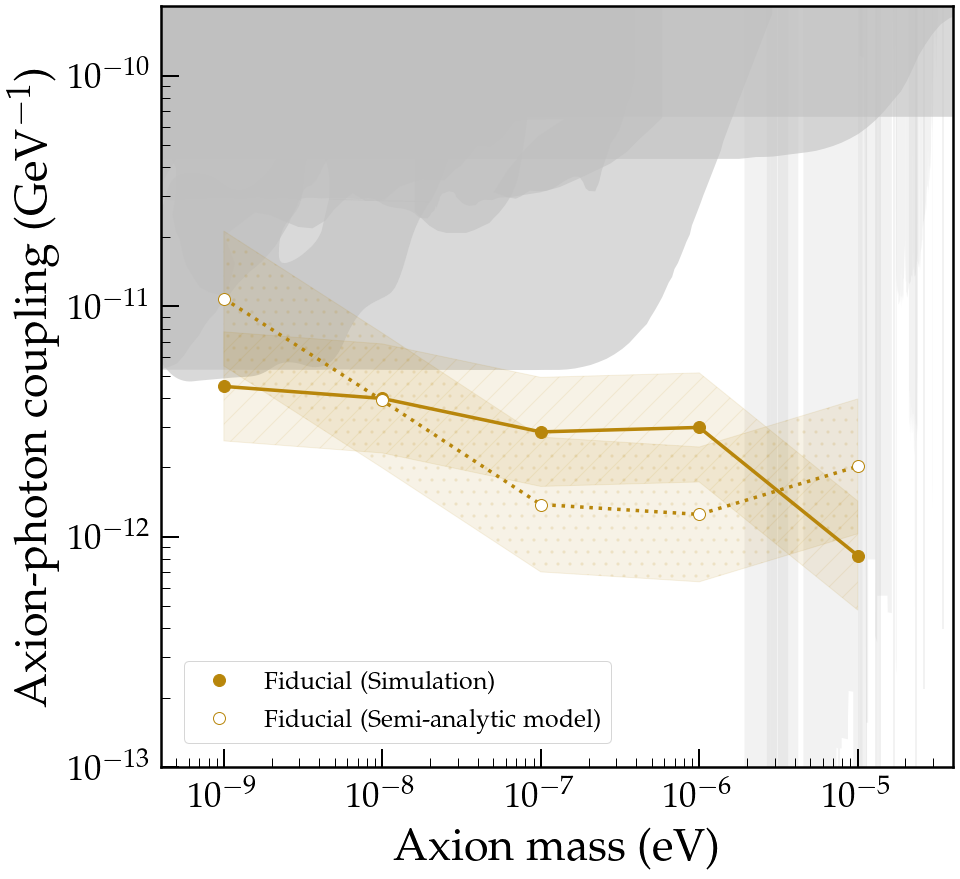}
    \caption{Fiducial upper limits derived for the simulation (solid) and semi-analytic model (dotted), shown together with corresponding bands depicting systematic model uncertainty.}
    \label{fig:sa_sim}
\end{figure*}

\begin{figure*}
    \centering
    \includegraphics[width=.48\textwidth]{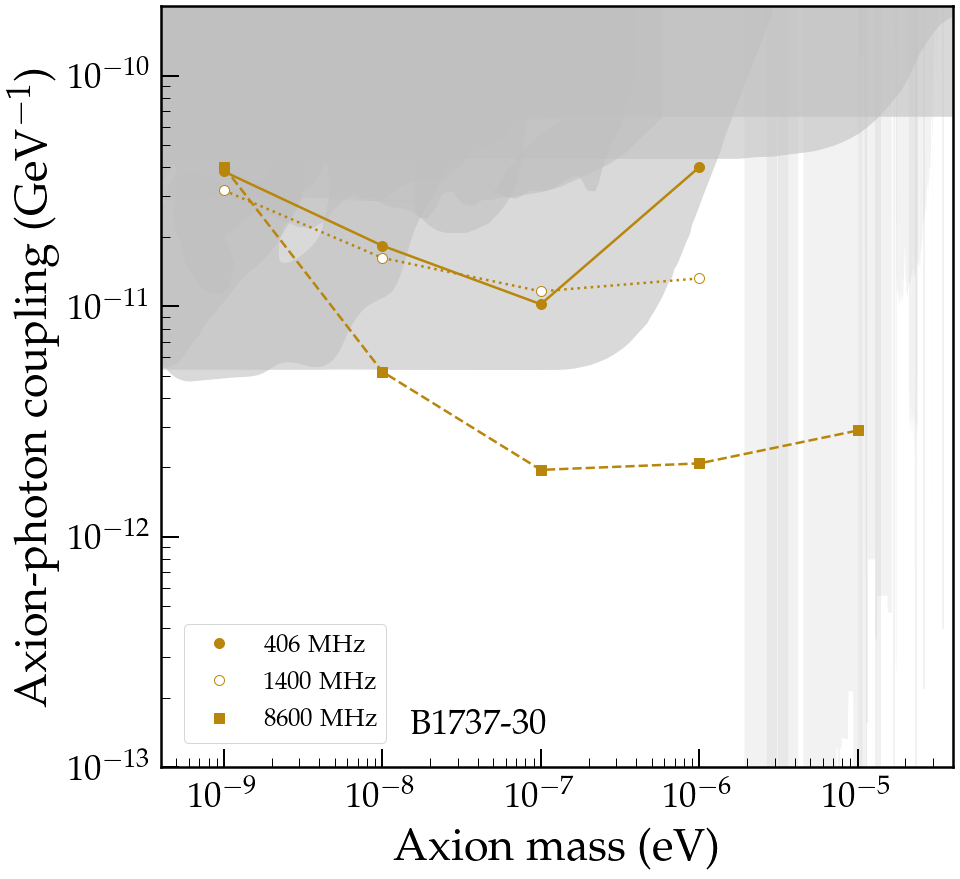}
    \includegraphics[width=.48\textwidth]{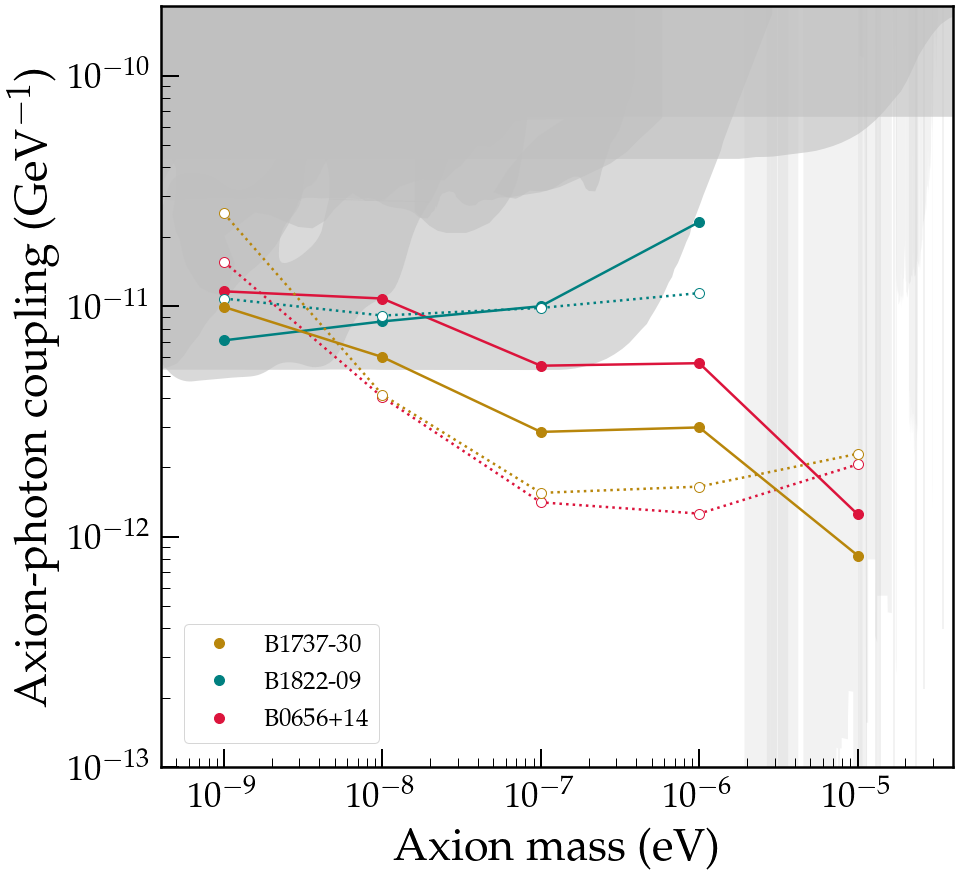}
    \caption{Left: Upper limits derived from one of the strongest pulsars (in the semi-analytic model) using exclusively observations of low ($\sim 408$ MHz), mid ($\sim 1.4$ GHz), or high ($\sim 8.6$ GHz) frequencies. Right: Comparison of the limits derived from three of the strongest pulsars using both the simulation (solid) and the semi-analytic model (dotted). The bounds for each pulsar differ by no more than a factor of a few between models. }
    \label{fig:bound_check}
\end{figure*}

The derived limits for the simulation and semi-analytic model are shown in Fig.~\ref{fig:sa_sim}, where the lines represent the fiducial constraints and the bands reflect the uncertainty due to particular modeling choices - the bands are also shown in Fig.~\ref{fig:Limits} and discussed thoroughly in the next section. We find that the final limits obtained through our pipeline are largely driven by observations of a few pulsars at higher frequencies. The reason for the latter is that the pulsar flux tends to drop much faster than the predicted axion flux with increasing frequency. In order to illustrate the origin of the constraining power, we plot in Fig.~\ref{fig:bound_check} the constraints derived from one of the strongest pulsars (in the semi-analytic model) using exclusively observations of low ($\sim 408 \, \rm MHz$), mid ($\sim 1.4 \, \rm GHz$), or high ($\sim 8.6 \, \rm GHz$) frequencies (left panel). We also plot in the right panel of Fig.~\ref{fig:bound_check} the individual constraints obtained from  both models using the three strongest pulsars (utilizing data at all frequencies). We find that the relative differences between the two models is typically no larger than a factor of $\sim 2$ in coupling (although larger differences appear for a small amount of outliers, like pulsar B0656+14 at intermediate masses), providing confidence that reasonable changes to our axion production model have a minimal impact on the final constraints. 

\section{Uncertainties}\label{sec:uncert}
This section reports the various uncertainties that enter our calculations, including uncertainties associated with the inferred pulsar properties, uncertainties in treating the PIC simulation, and uncertainties associated with the free parameters entering the semi-analytic model. We also provide quantitative estimates of the effects due to these uncertainties on the final constraints.

\subsection{Observational uncertainties}
Let us begin by discussing uncertainties associated to the inferred properties of each pulsar. In general, the properties that are measured to high precision for every pulsar are the rotational period $P_{\rm NS}$ and the spin-down rate $\dot{P}_{\rm NS}$. The value of the surface dipolar magnetic field $B_0$ can be inferred from these quantities. This can be seen by comparing the net Poynting flux through the light cylinder of a misaligned rotator, given by $L = \mu^2 \Omega_{\rm NS}^4(1 + \sin^2\theta_m)$ (where $\mu = B_0 \rns^3$, $\Omega_{\rm NS} = 2\pi/P_{\rm NS}$, and $\theta_m$ is the misalignment angle~\cite{LiSpitkovsky2012}), to the spin-down luminosity, yielding
\begin{align} \label{eqn:B0}
    B_0 = \frac{1}{2\pi} \left(\frac{I_{\rm NS}}{R_{\rm NS}^6 (1 + \sin^2\theta_m)} \right)^{1/2} \left(P_{\rm NS} \dot{P}_{\rm NS} \right)^{1/2}.
\end{align}
There are two points worth noting here. First, the ATNF adopts a fiducial moment of inertia $I_{\rm NS} = 10^{45} \, {\rm g \, cm^2}$, which is actually 30-50$\%$ lower than current estimates~\cite{steiner2015using}.
Second, the ATNF inferred field strengths assume $\theta_m =0$. One can see that these effects somewhat offset, however likely lead to a slight underestimation of $B_0$ at the level of $\sim \mathcal{O}(10\%)$. 

Another important consideration is that Eq.~\ref{eqn:B0} only reflects the magnitude of the dipolar magnetic field -- near the surface of the neutron star additional higher order multipoles could, and in some cases are expected to, contribute. This likely implies that the values of $B_0$ listed in Table~\ref{tab:pulsar} underestimate the true surface magnetic field strength (and thus we are likely underestimating our signal, potentially by a significant amount). Incorporating higher multipoles, however, must be done self-consistently, and thus in what follows we adopt the assumption that the magnetic fields are purely dipolar. We believe that this is likely a conservative assumption, however we plan to address this more rigorously in future work.

Since the derived limit is expected to scale like $g_{a\gamma\gamma}^{\rm lim} \propto B_0^\alpha$ with $1 \lesssim \alpha \lesssim 1.5$ (depending on the efficiency of the resonance), we adopt in what follows a characteristic uncertainty on $g_{a\gamma\gamma}^{\rm lim}$ of $\pm 20\%$. We expect this to be conservative, as it underestimates the impact of the corrected moment of inertia and neglects the contribution of higher order multipoles, both of which are expected to enhance the signal.

Notice that in estimating the uncertainty in the dipolar field strength we have neglected the contribution coming from the unknown equation of state, which allows for $\sim 10\%$ fluctuations in $R_{\rm NS}$~\cite{Lattimer:2021emm}. This is because the neutron star radius enters at multiple points in the analysis, and thus we treat this as an independent uncertainty. In particular, we expect the neutron star radius to enter with the following scalings: $r_{\rm pc}\propto R_{\rm NS}^{3/2}$, $E_{||} \propto r_{\rm pc} \propto R_{\rm NS}^{3/2}$, and $B_0 \propto R_{\rm NS}^{-3}$. Roughly speaking, this translates into a scaling on the limits as $g_{a\gamma\gamma}^{\rm lim} \propto (B_0^{4\alpha} \, E_{||}^2 \, r_{\rm pc}^4 )^{-1/4} \propto R_{\rm NS}^{9/4}$, where as before $1 \lesssim \alpha \lesssim 1.5$. Thus, a $10\%$ variation in the neutron star radius amounts to a $20\%$ shift in the limits. 

There exist a variety of techniques for inferring the distances to observed pulsars. By far the most reliable of these are parallax measurements, but unfortunately such measurements are only available for a small number of known pulsars. One can also analyze the HI spectrum on- and off-pulse, and use this information to infer the characteristic rotational velocities of the HI regions behind- and in front of the pulsar. In many cases, however, this information is not available. Without additional information, one must rely on inferring the distance using the observed dispersion measure (${\rm DM} = \int d\ell \, n_e$), which requires a model for the Galactic electron density. In Table~\ref{tab:pulsar} we compile inferred distances and uncertainties using various combinations of the techniques listed above. Priority is given to geometric measurements, and we only rely on dispersion measure distances when no other information is available (as done in the ATNF)~\cite{Verbiest:2012kh,yao2017new,Brownsberger:2014yya,price2021comparison}. In some cases, no published distance could be found; here, we used a variety of Galactic electron density models to find the distances using the tools available in~\cite{price2021comparison}. In a few cases, uncertainties on the distance were not readily available. In these cases, we compared the inferred distance from two different Galactic electron density models -- if the models showed reasonable agreement (differing by no more than a factor of $2$), we inferred the standard deviation from these measurements. In all other cases we defaulted to the more recent model~\cite{yao2017new} and placed a $100\%$ uncertainty on the distance. It is worth emphasizing that none of these pulsars (denoted in Table~\ref{tab:pulsar} with a $\dagger$) contribute meaningfully to the derived constraints. 

Finally, let us mention that the uncertainties on the pulsar geometry (the misalignment angle and orientation) as interpreted within the rotating vector model (RVM)~\cite{radhakrishnan1969magnetic} for the pulsars used in this analysis are expected to be small (see \eg~\cite{gupta2003understanding}). Note that this is not a generic feature of all pulsars, however we have chosen our pulsars largely to minimize this uncertainty. There is of course a larger systematic uncertainty that pertains to the validity of the model itself, but it is unclear how this can be quantified. Given that the RVM is the state-of-the-art, we take these values to be fixed in our analysis -- future analyses which are applied over large populations of pulsars may be able to evade the need for specific assumptions on pulsar geometry. We leave such an endeavour for future work.

\subsection{PIC Simulation}
There are two clear identifiable uncertainties that enter our analysis of the PIC simulation. First, the physical dimensions of the simulation are expressed in terms of the polar-cap radius. In general, the fractional surface area which leads to pair cascades depends on a variety of factors including the magnetic field geometry, the misalignment angle, the properties of the return current, the compactness of the neutron star, and various nonlinearities in the electrodynamics. While it is difficult to quantify the effects of each of these separately, it is expected that together they can lead to an $\mathcal{O}(1)$ modification of the polar-cap radius~\cite{TimokhinArons2013,beloborodov2008polar,belyaev2016spatial}\footnote{Note that pulsars with extremely large non-dipole multipoles can have smaller polar-cap radii. This however comes at the expense of an enormous magnetic field, which is expected to entirely offset the resultant decrease in axion production~\cite{szary2015two}.}. Axion production at low masses scales with the square of the volume (with a weakening of the dependence on the volume at higher masses), meaning we expect our limit on the axion photon coupling to scale proportional to this uncertainty. In what follows we therefore adopt a characteristic  uncertainty on $r_{\rm pc}$ at the level of $\pm 30\%$\footnote{Note that the PIC simulation intrinsically assumes an aligned rotator (oblique rotators would require going to 3+1D simulations). The physics of pair production and screening is expected to be same for oblique rotators, with the dominant effect being a shift in the geometry and the potential drop -- this, however, is an implicit assumption which must be verified in the future.}, which roughly corresponds to the one-sigma range inferred from a flat distribution varying between $\pm 50\%$ of the fiducial value. 

The second source of uncertainty comes from the re-scaling factor $\xi$, which is intended to undo the compression of scales that is performed in the simulation. Unfortunately, without more simulations at various resolutions it is impossible to truly understand the impact of this re-scaling factor. Given, however, that the semi-analytic model also contains a slight scale compression (and subsequent re-scaling), we adopt the uncertainty inferred from the semi-analytic model in the simulation as well (see following subsection). We do note that this procedure is not exactly valid given that the scale compression is not done in an equivalent manner -- it is largely for this reason that the semi-analytic model was constructed in the first place.

Finally, it is worth mentioning that there also exists an unquantifiable uncertainty that enters the analysis of the PIC simulation, which arises from the fact that the physical parameters used in the simulation do not directly correspond to those of true pulsars (recall that this is because the parameter re-scalings performed in the simulation modify the fundamental scales of the problem). In the future, one would ideally like to perform a number of simulations varying the re-scaling parameters in order to understand the sensitivity of axion production to this procedure, however at the moment this is computationally unfeasible. Thus, we instead use the semi-analytic model as a cross-check, allowing us to ensure the robustness of the results.

\begin{figure*}
    \centering
    \includegraphics[width=.48\textwidth]{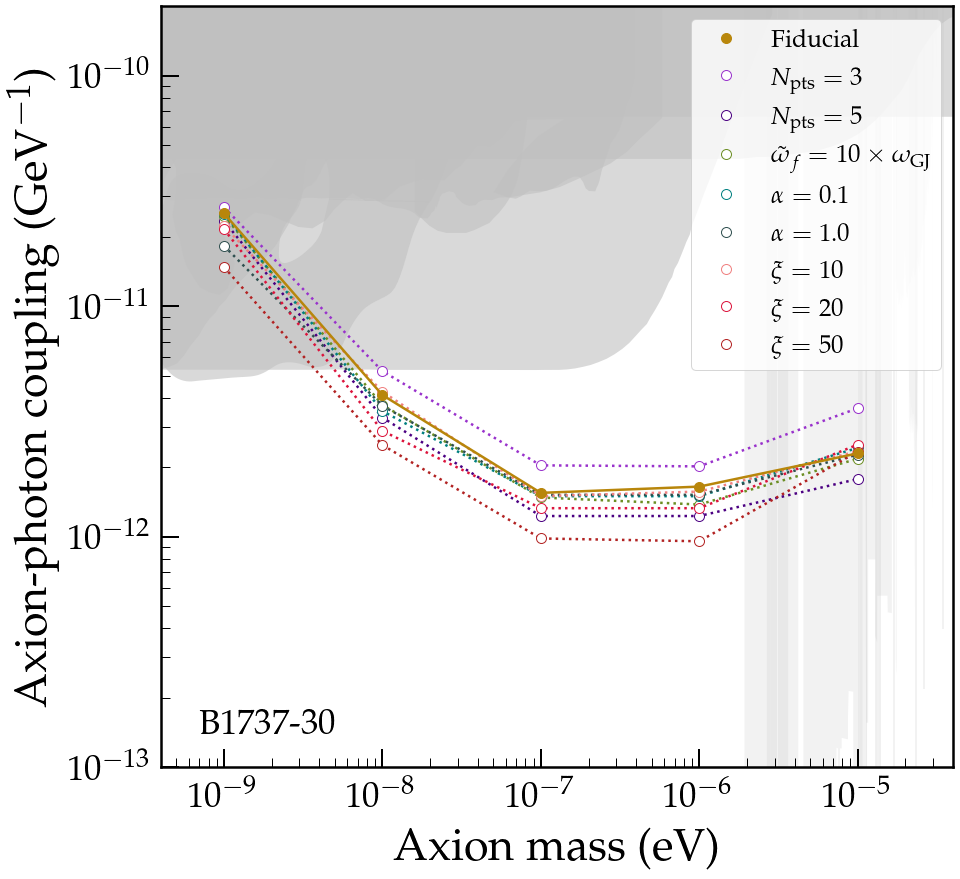}
    \caption{Upper limits derived using observations of B1737-30 and the semi-analytic model for the axion production rate. We asses the impact of varying several unconstrained parameters (namely $N_{\rm pts}$, the final value of $\tilde{\omega}$, $\alpha$, and $\xi$) on the derived limits, and compare with the fiducial model (solid, gold). }
    \label{fig:sa_system}
\end{figure*}

\begin{table*}
\setlength\extrarowheight{5pt}
    \centering
    \begin{tabular}{|c|c|c|c|c|} \hline
         Parameter & Characteristic uncertainty on $g_{a\gamma\gamma}^{\rm lim}$  &  Simulation & Semi-analytic  \\  \hline  \hline
         $B_{\rm dipolar}$ & $ \pm 20\%$ &  $\checkmark$ & $\checkmark$ \\ \hline
         $R_{\rm NS}$ & $ \pm 20\%$ & $\checkmark$ & $\checkmark$ \\ \hline
         $r_{\rm pc}$ & $ \pm 30\%$ & $\checkmark$ & $\checkmark$ \\ \hline
         $N_{\rm pts}$ & $ \pm 25\%$ &   $\tikzxmark$ & $\checkmark$ \\ \hline
         $\hat{\omega}_f$ & $ \pm 5\%$ &  $\tikzxmark$ & $\checkmark$ \\ \hline
         $\alpha$ &  $ \lesssim \pm5 \%$  &  $\tikzxmark$ & $\checkmark$ \\\hline
         $\xi$  & $\lesssim \pm 5\%$ 
 &   $\checkmark$ & $\checkmark$ \\ \hline
         $\epsilon$ & $\sim 0\%$ &    $\tikzxmark$ & $\checkmark$ \\ \hline
         $\kappa$ & $\sim 0\%$ &   $\checkmark$ & $\checkmark$ \\ \hline \hline
         Total &  & $\pm 42 \%$ & $\pm 49 \%$ \\  \hline
    \end{tabular}
    \caption{ Estimated breakdown of the characteristic uncertainty on the derived limits arising from  systematic uncertainties in the semi-analytic model and simulation (we denote in each column whether the listed uncertainty is included and relevant for each model). The total uncertainty is obtained by combining the individual contributions in quadrature, and is reflected in the final row of each model.  }
    \label{tab:uncert}
\end{table*}

\subsection{Semi-analytic Model}
As with the simulation, the semi-analytic model is subject to uncertainties in the characteristic polar-cap size. Additionally, the semi-analytic model has a number of free parameters which have been fixed to fiducial values in derivation of the upper limits. Here, we outline the impact of reasonable variations in each of these. The primary free parameters include: $N_{\rm pts}$ (the number of seed points used in the plane wave damping), $\hat{\omega}_f$ (the final plasma frequency at the end of the particle cascade), $\alpha$ (the power law index controlling how quickly the plasma frequency evolves after the initial collapse), $\xi$ (the re-scaling factor), $\epsilon$ (the parameter controlling the acceleration time of particles in the gap), and $\kappa$ (the parameter describing the initial plasma density present in the gap prior to the pair cascade, note that this parameter also enters the simulation).

The latter two parameters (namely, $\epsilon$ and $\kappa$) have no perceptible effect on the final results. In the semi-analytic model, the acceleration time (see Eq.~\eqref{eqn:tacc}) is the smallest length-scale in the problem. This means that for the fields considered in this work, the acceleration time is often smaller than the time resolution of the model. As a result, we generally accelerate each particle to its final Lorentz factor within a single time step. The value of $\epsilon$, which describes at what fraction of the radiation-reaction limited Lorentz factor each particle begins to emit CR photons, shows up logarithmically in the acceleration time and thus has at most an $\mathcal{O}(1)$ effect on $\tacc$, which is already small compared to the time resolution. Therefore, even orders of magnitude changes in $\epsilon$ have negligible effect on the final result. The parameter $\kappa$ is introduced to avoid having to simulate $\sim n_{\rm GJ} \rpc^3$ particles, which would be computationally infeasible. Since the number density grows exponentially during a cascade, $\kappa$ can be understood as a logarithmic re-scaling of the time in the pair cascade process. However, we note that the pair cascade process is introduced to simply set the initial conditions in the gap prior to dynamical screening. The initial conditions and the dynamics of the screening phase are independent of $\kappa$, and thus the effect of $\kappa$ on the final result is also negligible.

In order to assess the sensitivity of the presented limits to the choice of the other parameters we run an independent analysis on one of the most-constraining pulsars, B1737-30. In particular, we perform an extra run with $N_{\rm pts} = 3$, one with $N_{\rm pts} = 5$ (rather than $N_{\rm pts} = 4$),  a run in which $\hat{\omega}_f = 10$ (rather than $100$), two runs in which $\alpha$ is varied from 0.1 to 1.0, and various runs using a larger stretching parameter $\xi$. The limits obtained in each of these runs are displayed alongside our fiducial limit from this pulsar in Fig.~\ref{fig:sa_system}. The characteristic uncertainty of each parameter variation on the limit is quantified and included in Table~\ref{tab:uncert}. Note that in the special case of $\xi$, we infer the uncertainty by looking at the limit in which $\xi \rightarrow 1$ rather than by comparing variations across runs.

Most of the uncertainties listed in Table~\ref{tab:uncert} are extremely sub-dominant to that coming from the uncertainty associated to the polar-cap size, with the variation in $N_{\rm pts}$ otherwise producing the largest effect. The total uncertainty for each model (estimated at the level of $36\%$ and $45\%$ for the simulation and semi-analytic models, respectively) is obtained by summing the errors in quadrature, and is reflected in a band about the fiducial limits in Fig.~\ref{fig:Limits} and Fig.~\ref{fig:sa_sim}.

\end{document}